\documentclass[a4paper,11pt]{article}
\pdfoutput=1 
\usepackage{jcappub}
\usepackage[T1]{fontenc} % if needed
\usepackage{xcolor}

\title{\boldmath Inclusive fluxes of secondary air-shower particles}

\author{Hariom Sogarwal}
\author{and Prashant Shukla}
\affiliation{Nuclear Physics Division, Bhabha Atomic Research Centre, Mumbai 400085, India} 
\affiliation{Homi Bhabha National Institute, Anushakti Nagar, Mumbai 400094, India}

\emailAdd{sogarwalhariom@gmail.com}
\emailAdd{pshuklabarc@gmail.com}

\abstract{
The particle showers produced in the atmosphere due to the interactions of primary 
cosmic particles require a thorough understanding in the backdrop of searches for
rare interactions. 
In this work, we made a comprehensive study of air shower simulations using various
combinations of hadronic models and particle transport code of the CORSIKA package.
The primary proton and helium distributions are taken as power law 
which are scaled to match the measured flux in balloon and satellite-based experiments at the top of 
atmosphere. The shower simulation includes production, transport and decays of 
secondaries up to the ground level.
In this study, we focus on the bulk of the spectra and particles which is 
computationally intensive and hence parallel processing of events is done on
computer cluster. 
We provide a way to normalize the simulation results to be  compared with
the ground-based measurements namely, single and multiple muon yields and
their charge ratios as a function of zenith angle and momentum. This provides
a basis for comparisons among the six model combinations used in this study
and the differences are outlined.
Most of the hadronic models in CORSIKA produce the bulk ground based
measurements fairly well.
We use one of the best model combinations to quantitatively predict the
absolute and relative yields of various particles at ground level as well as
their correlations with primaries and with each other. The leptonic ratios
are obtained as a function of energy and zenith angle which are important
inputs for the neutrino oscillation physics. 
}

\begin{document}

\maketitle
\flushbottom

\section{Introduction}
\label{sec:intro}

The cosmic rays consisting of high energy charged particles continuously
bombard the Earth from all directions and have remained one of the most
interesting and active areas of research~\cite{grieder444507108cosmic}.
The particle showers produced in the atmosphere due to the interactions of
these primary cosmic particles have provided a natural laboratory of physics
of standard and beyond standard models which led to many discoveries.
Accurate knowledge of both the primary as well as secondary particles is
an important prerequisite for search of rare interactions.  
The primary cosmic rays consist of protons (90\%), helium nuclei (9\%) 
and other heavy nuclei (1\%) which have been produced and accelerated
in astrophysical processes to high energies as large as $10^8$ TeV.
Upon entering the atmosphere, the primary cosmic particles interact with the 
nuclei of oxygen and nitrogen gases roughly 12 km above Sea level. 
These interactions lead to the production of shower of hadrons, mainly 
pions and kaons with a small percentage of heavier particles which move
downwards while losing energy and decaying.
These particles introduce cosmogenic signal and 
background for the measurement of neutrino oscillation physics and hence their 
study is very useful to precisely understand the mass difference of the neutrinos 
in the 2-3 sector and the mass hierarchy.
Although neutrinos are the most abundant shower particles at the Earth surface,
muons are the most abundant charged particles and are responsible for half
of the terrestrial radiation dose taken by a body on Earth.
Measurements of muon distributions at ground is an active
field and the improved theoretical understanding and detection technologies are
providing more insights in the particle production in showers
~\cite{Sogarwal:2022kgw,ParticleDataGroup:2022pth}.
LHC is working on adedicated experiment of proton-oxygen 
collisions to understand the air shower physics.
This dedicated LHC run will help in tuning of various hadronic interaction 
parameters used in the cosmic ray packages for better understanding of the 
air shower produced in the atmosphere and spectra of various particles at ground level
~\cite{Dembinski:2020dam}.

 The shower simulation codes employ hadronic interaction models which have
energy dependent hadronic cross-section calculations as the main body of 
the simulation. CORSIKA is a Monte Carlo program for detailed simulation 
of air showers initiated by high-energy primary cosmic particles such as
protons~\cite{Heck:1998vt} using various hadronic interaction models which 
can be selected in the CORSIKA options (version 7.7100)~\cite{userguide}.
For the hadronic interactions at high energies, the most popular models are
QGSJET and SIBYLL. QGSJET~\cite{QGS1,QGS2,QGS3} (Quark Gluon String model with JETs)
describes high-energy hadronic interactions using the quasi-eikonal pomeron 
parameterization for the elastic hadron-nucleon scattering amplitude.
The hadronization process is treated in the quark gluon string model. 
The more improved version is QGSJET-II-04~\cite{QGSII,QGSII1} which 
includes pomeron loop and the cross-sections are tuned to LHC data.
SIBYLL~\cite{fletcher1994s,sibyll,ahn2009cosmic,riehn2015hadronic,riehn2015new}
is a program developed to simulate hadronic interactions at extremely high 
energies based on the QCD mini-jet model.
SIBYLL also activates the inelastic hadronic interaction cross-sections at
higher energies which are based on QCD calculations.
We use SIBYLL 2.3c version~\cite{riehn2015hadronic}.

For low-energy (below 100 GeV) interactions, the popular packages are GHEISHA,
UrQMD and FLUKA. GHEISHA (Gamma Hadron Electron Interaction SHower code) 
is a model which describes hadronic collisions up to some 100 GeV in many 
experiments. A detailed description of the physics processes covered by GHEISHA 
may be found in Ref.~\cite{Gheisha}.
UrQMD (Ultra-relativistic Quantum Molecular Dynamics) is another package
designed to treat low-energy hadron-nucleus 
interactions. A detailed description of this model may be
found in Ref.~\cite{bass1998microscopic,bleicher1999relativistic}. 
UrQMD 1.3 is used in CORSIKA to perform the elastic and inelastic interactions 
of hadrons below 80 GeV in air.
FLUKA (FLUctuating KAscade)~\cite{Bohlen:2014buj,Ferrari:2005zk} is a package of
routines which transport energetic particles through matter by the monte carlo method.
In combination with CORSIKA, only the low-energy hadronic interaction part is used which 
calculates the inelastic hadron cross-sections with the components of air and 
produce secondary particles including many details of the de-excitation of the
target nucleus.
 CORSIKA is also used for generating showers for small local
arrays~\cite{Wibig:2021pim}. There are also plans 
concerning the design and development for the new generation of 
CORSIKA for various scientific applications making an efficient use of 
computational resources~\cite{Engel:2018akg}.

CORSIKA is widely used for generating muons for detector
simulations~\cite{Barber:2017rvr,Mueller:2015thh}. 
There have been several studies to compare and validate different CORSIKA models
with experimental data.
In one of the initial studies~\cite{Wentz:2003bp}, muon and neutrino
flux ratios were obtained using various models of CORSIKA and comparison
was made with experimental data and other theory models. 
The work in Ref.~\cite{Djemil:2005hr} used CORSIKA to obtain the muon
momentum distributions at ground and at an altitude using VENUS and two
low energy models GHEISHA and UrQMD which are compared with the measurements.
Recently, work in Ref.~\cite{Nikolaenko:2021oyi} used four high energy models
of hadronic interactions in CORSIKA to compare the distributions of different
particles. The study outlines the differences between different models.
The work in Refs.~\cite{Halataei:2008zz,Patgiri:2016odq} used CORSIKA to
study the variations of shower angle distribution as a function of atmospheric depth
from tens of TeV to PeV energies of protons. 

There have been few recent studies of muon observables obtained with CORSIKA.
The CORSIKA is used to study the dependence of the power index of angular distribution
of muons on the energy and atmospheric depth using the QGSJET and
GHEISHA models~\cite{Bahmanabadi:2018dbr}.
Bahmanabadi~\cite{Bahmanabadi:2019wdx} measured the muon charge ratio in the low
energy range and made a comparison with two low energy models (GHEISHA and UrQMD)
in CORSIKA.
Bahmanabadi and Fazlalizadeh ~\cite{Bahmanabadi:2019kel} 
studied the east-west asymmetry as well as the muon charge ratio 
sensitivity to the low-energy GHEISHA and UrQMD in combination with the high 
energy QGSJET-II model.
In Ref.~\cite{Cohu:2021tjh}, the momentum and angle distributions
of muons obtained using CORSIKA are compared with the analytical formulae.

%In 2022, Wibig~\cite{Wibig:2021qkp} studied 
%the superposition model in CORSIKA for primary energies smaller than 
%$<$ 10$^{15}$ eV in air shower simulations.
%The work in Ref.~\cite{Hariharan:2019xgn} discussed the method which
%reduces the computation time by cutting down the primary particles
%below the rigidity cutoff Rc which has been incorporated in
%CORSIKA version v75600 onwards.

Many of the studies in the literature focussed on the hadronic models for very high-energy
showers. Also many studies concentrated on the ratios of the particles
rather than absolute yields. 
We planned our work to make a comprehensive study of air shower simulations
using different combinations of low and high energy models in CORSIKA.
In this study, we focus on the bulk of the particles at ground level
and obtain the absolute yields using normalized simulation results.
We have used data for muon distributions namely, zenith angle distribution,  
momentum distributions (at $0^\circ$ and $75^\circ$) and charge ratios
to compare with six model combinations. 
The predictions are made for multiplicity and East-west asymmetry for muons
for all model combinations. The leptonic distributions and ratios form
an important input for neutrino oscillation studies. 

Contribution in the different muon observables coming from different energy as
well as zenith angle intervals of primary protons and heliums is obtained to
correlate the ground based measurements with the initial particle direction and energy. 
We predict the zenith angle and momentum distributions of
all other particles and their ratios which could be measurable at ground
in addition to muons. The distributions of other hadronic and leptonic particles
are obtained. 
Distributions of muons and neutrinos as a function of $\theta$ and $\phi$
is presented to quantify East-west asymmetry.

We perform CORSIKA simulation using various hadronic interaction models and 
obtain various distributions of particles at the ground.
The low-energy hadronic interaction models used in this work are GHEISHA, UrQMD and FLUKA,
whereas the high-energy interaction models are QGSJET, QGSJET-II and SIBYLL.
The initial cosmic proton and helium distributions have been taken as power law 
which are matched with experimental data from balloon and satellite-based measurements.
The muon distributions obtained are then compared with the experimental data
measured at the ground level and with some of the recent parametrizations. 
In view of a dedicated run of proton oxygen collisions planned at LHC, such
study becomes more important and timely.

\section{Parameterizations of muon distributions}  
In this section, we describe the parametrizations which are used to compare
with the CORSIKA simulations. 
The energy distribution of primary cosmic rays follows power law $E^{-n}$. 
The pion and the muon distributions also follow the same power law which 
is modified in the low-energy region~\cite{shukla2016energy} as

\begin{eqnarray}
I(E)= I_{\circ} N (E_{\circ} + E)^{-n}\bigg(1+\frac{E}{\epsilon}\bigg)^{-1}.
\label{eq:shukla}
\end{eqnarray}
Here, $N= (n-1)\,(E_{\circ}+E_{c})^{(n-1)}$ is the normalization in which $E_{c}$ (= 0.5 GeV) is the cut-off value of the data while and $E_{\circ}$ and
$\epsilon$ are parameters. 
The zenith angle distribution of energy integrated flux in terms of
$I_{\circ} = \Phi(\theta = 0)$ is obtained as

\begin{eqnarray}
\Phi(\theta) = I_{\circ}\ D(\theta)^{n-1},
\label{eq:ang}
\end{eqnarray}
where D($\theta$) is the ratio of pathlength of a muon from inclined 
direction to that of a muon from the vertical direction and is given 
by~\cite{shukla2016energy}

\begin{eqnarray}
D(\theta) = \sqrt{ \dfrac{R^2}{d^2}\cos^2\theta+2\dfrac{R}{d} + 1  }
           - \dfrac{R}{d}\cos\theta.
\end{eqnarray}
Here, $d$, $R$ and $\theta$ are the vertical depth, radius of the Earth and 
the zenith angle respectively. For the case when the Earth is assumed 
to be flat, $D(\theta)$ = $1/\cos\theta$, the Eq.~\ref{eq:ang} leads to

\begin{eqnarray}
  \Phi(\theta) = I_{\circ} \cos^{n-1}\theta.
  \label{eq:angular}
\end{eqnarray}
With $n \simeq 3$, this gives the usual $\cos^2\theta$ distribution 
which is empirically known to describe zenith angle distribution.
The values of the parameters for muon energy distributions corresponding to Eq.~\ref{eq:shukla} are listed
in the Table~\ref{table1}~\cite{shukla2016energy}. Table~\ref{table2} lists the values of parameters
of muon zenith angle distribution corresponding to Eqs.~\ref{eq:ang}
and \ref{eq:angular}~\cite{shukla2016energy}. 
Table~\ref{table3} lists the values of parameters of proton and helium energy distributions
corresponding to Eqs.~\ref{eq:shukla} which have been obtained in the present work.

\begin{table*}[h]
\begin{center}
\caption{Parameters of Eq.~\ref{eq:shukla} for muon energy distributions~\cite{shukla2016energy}.}
\label{table1}
\begin{tabular}{|c| c| c| c| c|} 
\hline
&  $I_{\circ}$	       & $n$          & $E_{\circ}$         & $1/\epsilon$    \\
& (m$^{-2}$s$^{-1}$sr$^{-1}$) &              & (GeV)         & (GeV)$^{-1}$     \\
\hline
$\mu$ at $0^{\circ}$ & 70.7$\pm$0.2      & 3.01$\pm$0.01 & 4.29$\pm$0.04  & 1/854      \\
Sea level (E$>$0.5 GeV)    &                     &               &                &           \\
\hline
$\mu$ at $75^{\circ}$	&  65.2$\pm$1.5         & 3.00$\pm$0.02 & 23.78$\pm$0.30  & 1/2000  \\
Sea level (E$>$1.0 GeV) &                     &               &                 &          \\
\hline
\end{tabular}
\end{center}
\end{table*}

\begin{table}[h]
\begin{center}
\caption{Parameters for zenith angle distribution functions of muons~\cite{shukla2016energy}.}
\label{table2}
\begin{tabular}{|c| c| c| c| c|} 
\hline	
Function                          & $I_{\circ}$         & $n$               & $R/d$      \\
\hline
$\Phi(\theta) = I_{\circ}D(\theta)^{n-1}$  & 88.0 $\pm$ 2.4  & 3.09 $\pm$ 0.03   & 174 $\pm$ 12   \\
\hline
$\Phi(\theta) = I_{\circ} \cos^{n-1}\theta$  & 85.6$ \pm$ 2.4 & 3.01 $\pm$ 0.03  & -    \\
\hline
\end{tabular}
\end{center}
\end{table}

\begin{table*}[h]
\begin{center}
\caption{Parameters of Eq.~\ref{eq:shukla} for proton and helium energy distributions.}
\label{table3}
\begin{tabular}{|c| c| c| c| c| c|} 
\hline
&  $I_{\circ}N$	       & $n$          & $E_{\circ}$         & $1/\epsilon$  & $\chi^{2}$/NDF \\
&                      &              & (GeV)         & (GeV)$^{-1}$     & \\
\hline
Proton (E$>$1 GeV) & 16200$\pm$513   &  2.760$\pm$0.032  & 0.52$\pm$0.11     & 0.0 &  50.19 \\
\hline
Helium (E$>$1 GeV) & 759$\pm$292   &  2.703$\pm$0.002  & 0.31$\pm$0.55     & 0.0 & 34.57  \\
\hline
\end{tabular}
\end{center}
\end{table*}

Gaisser had given analytical formula~\cite{gaisser2002semi} for muon energy 
distribution assuming flat Earth and which is valid for high-energy 
$E_\mu > 100/\cos\theta$ GeV, is given by

\begin{eqnarray}
{\frac {dN_\mu} {dE. d\Omega}} & \approx 1400 E^{-2.7}_{\mu} / ({\rm m^{2}s GeV sr}) \Bigg( \frac {1} {1+ \frac {1.1 E \cos\theta} { \epsilon _\pi }}
 + \frac {0.054}  {1 + \frac{1.1 E \cos\theta} {\epsilon _ \kappa}} \Bigg), \nonumber
\label{eq:gaisser} 
\end{eqnarray}
where the two terms in the bracket give the contributions of pions and kaons in
terms of the two parameters $\epsilon_\pi \approx$ 115 GeV and
$\epsilon_\kappa \approx $ 850 GeV.

\section{CORSIKA simulation inputs}

In our simulations, the atmospheric model was chosen as the Central European
atmosphere for October month out of the available atmospheric 
models in the CORSIKA simulation.
%We have used the vertical muon flux measured by Haino et al.~\cite{haino2004measurements} 
%at Tsukuba during October 2002.
The conditions are taken as central Europe which are also a
close approximation to conditions at Tsukuba.
The horizontal component (towards the North) and  vertical component
(downwards) of the Earth's magnetic field are
taken as 30.07 $\mu$T and 35.32 $\mu$T respectively.
The observation level is a sphere following the Earth's curvature at
an altitude of 30 m.
%The muon multiple scattering angles is selected
%for large steps by Moliere's theory and for small steps by adding many
%single Coulomb scattering events. 

The primary proton/helium energy distribution is chosen as the power law of the form
\begin{eqnarray}
  I  = A \, E^{-2.7}.
\label{power27} 
\end{eqnarray}
The energy is generated in the range between 10 GeV and $5\times10^5$ GeV in
our simulations. The simulation process is computationally intensive hence
parallel processing of events at computer cluster has been performed. 
The primary particle is generated in nine intervals of energy given by 
[10, 50] GeV, [50, $10^2$] GeV, [$10^2$, $5\times10^2$] GeV, 
[$5\times10^2$, $10^3$] GeV, [$10^3$, $5\times10^3$] GeV, 
[$5\times10^3$, $10^4$] GeV, [$10^4$, $5\times10^4$] GeV, 
[$5\times10^4$, $10^5$] GeV and [$10^5$, $5\times10^5$] GeV. In each 
energy interval, 100000 events are generated except in the last interval
in which 1000 events are generated. 

The spectrum in each energy range is weighted by the factor $f$ given as
\begin{eqnarray}
f = \frac{E_1^{-1.7} - E_2^{-1.7}}{E_3^{-1.7} - E_4^{-1.7}},
\label{eq:1}   
\end{eqnarray}
where $E_1$ = 10 GeV, $E_2$= 50 GeV and $E_3$, $E_4$ are the lower and upper
bound of a given energy interval. With this method, then we obtain 
the equivalent number of proton showers generated over the whole distribution
in the energy range 10 - $5\times10^5$ GeV as "106,933".
%The measured 
%(integrated) number of protons and heliums at the top of the atmosphere 
%are 5003.05 $m^{-2}sec^{-1}sr^{-1}$ and 696.70 $m^{-2}sec^{-1}sr^{-1}$ 
%respectively.

The default primary intensity distribution of protons and heliums with the 
zenith angle $\theta$ in CORSIKA goes as
\begin{eqnarray}
I(\theta) \sim \, \sin\theta.
\label{eq:3}
\end{eqnarray}
The sine term respects the solid angle element of the sky. The $\phi$ distribution
is taken as uniform between $-\pi$ to $\pi$.
 
\begin{figure*}	
\begin{center}
\includegraphics[width=0.48\linewidth]{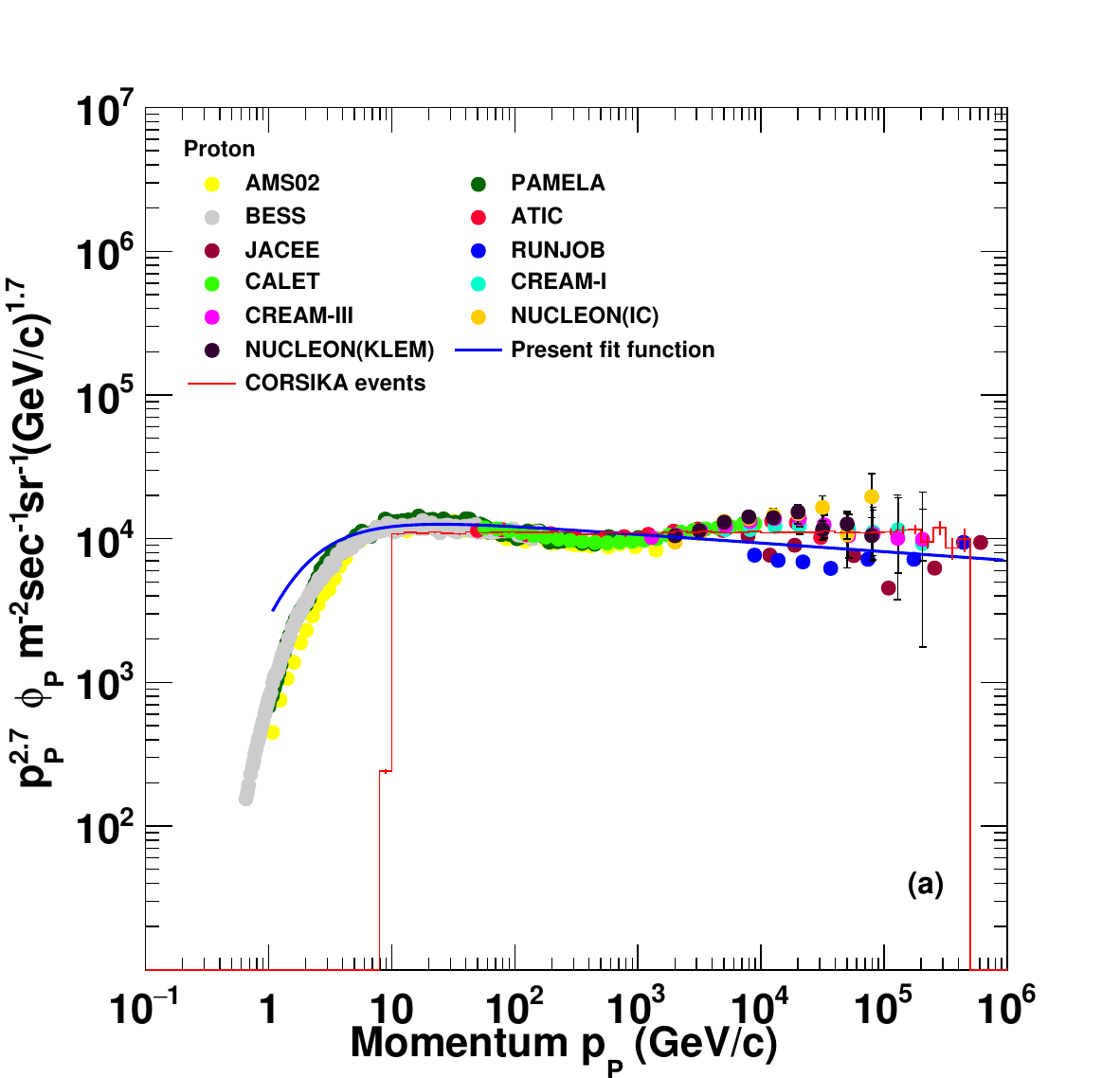}
\includegraphics[width=0.48\linewidth]{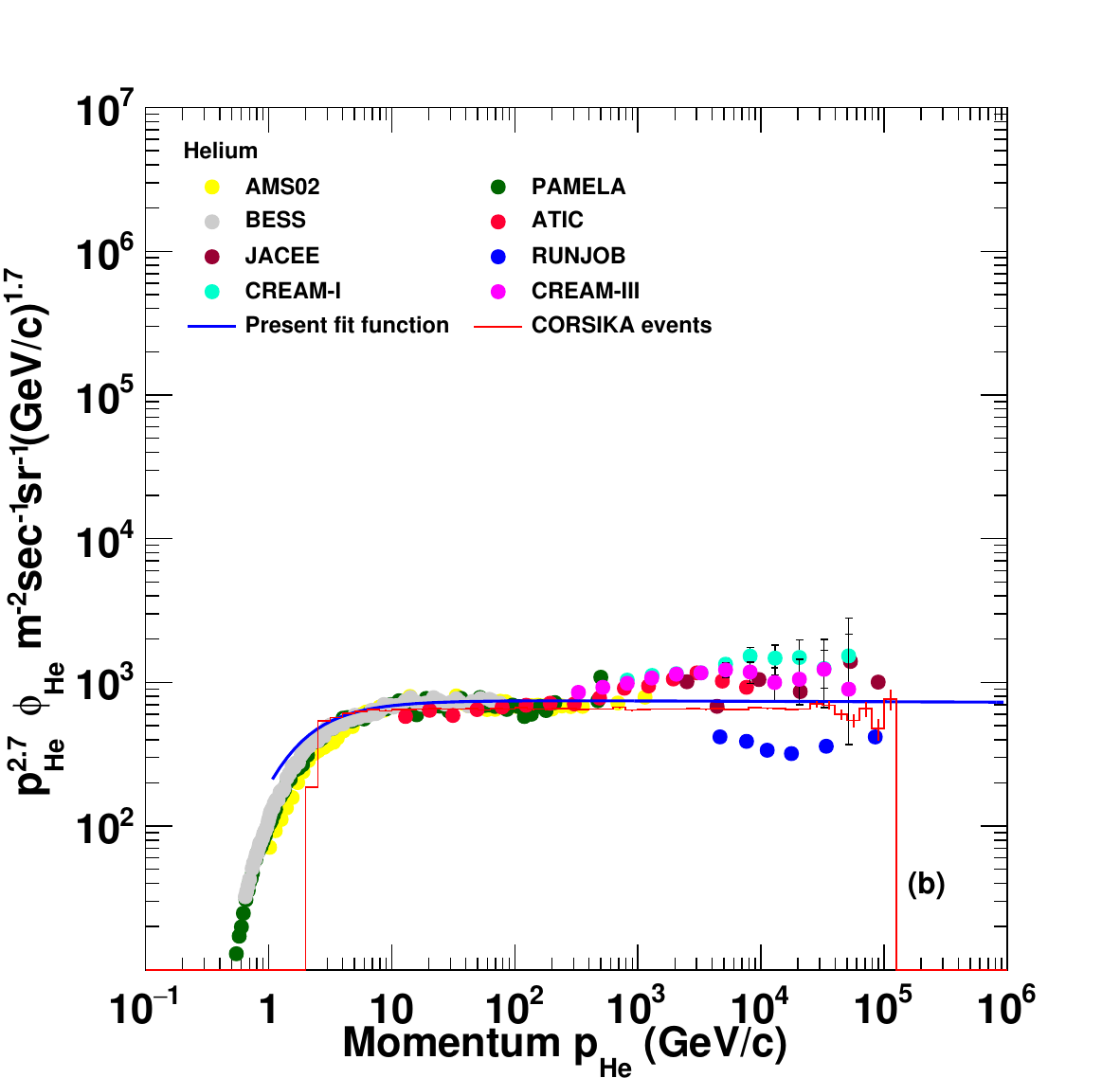}
\caption{(a) The proton flux as a function of momentum falling at
the top of the atmosphere as measured by balloon and satellite-based experiments 
~\cite{abe2016measurements, adriani2011, aguilar12015, asakimori1998, panov2009, derbina2005}
along with the initial proton spectrum used in the simulation and the
parameterization given by Eq.~\ref{eq:shukla}.
(b) The helium flux as a function of momentum per nucleon 
falling at the top of the atmosphere as measured by balloon and satellite-based experiments
~\cite{abe2016measurements, adriani2011, aguilar22015, asakimori1998, panov2009, derbina2005} 
along with the initial helium spectrum used in the simulation
and the parameterization given by Eq.~\ref{eq:shukla}.}
\label{figure1}
\end{center}
\end{figure*}

Figure~\ref{figure1}(a) shows the proton flux as a function of momentum  
falling at the top of the atmosphere as measured by balloon and satellite-based experiments
~\cite{abe2016measurements, adriani2011, aguilar12015, asakimori1998, panov2009, derbina2005}
along with the proton 
spectrum used in the simulation and the parameterization given by
Eq.~\ref{eq:shukla}. 
Figure~\ref{figure1}(b) shows the helium flux as a function of momentum 
per nucleon falling at the top of the atmosphere as measured by balloon and satellite-based  experiments~\cite{abe2016measurements, adriani2011, aguilar22015, asakimori1998, panov2009, derbina2005} 
long with the initial helium spectrum used in the simulation and the parameterization given by
Eq.~\ref{eq:shukla}.
The momentum distributions of initial protons and heliums used 
in the simulation are represented here by histograms having logarithmic 
bins.
It is observed that the momentum distribution of the primary particles roughly
follow power law ($p^{-2.7}$) in the energy range from 10 GeV/$c$ to $5\times10^5$ GeV/$c$
considered in the present work. 
The primary proton and helium spectra are matched to the measured spectrum 
and scale factors are obtained which are used to obtain normalization for all
the spectra obtained by protons and heliums in CORSIKA.
The normalized spectra obtained from protons and heliums are then added.

\section{Results \& discussions}		

In CORSIKA, among low energy models, we choose GHEISHA, UrQMD and FLUKA
and among high energy models we choose QGSJET, QGSJET-II and SIBYLL.
Both QGSJET  and QGSJET-II are the hadronic interactions models 
for the high-energies interactions and QGSJET-II is the upgraded form of QGSJET. 
Our interest was to see their similarity and differences in the bulk of the particles 
spectrum at ground level. In earlier investigations of cosmic muons at LEP, the hadronic 
interactions were simulated using QGSJET 01. Besides the treatment of nonlinear effects, 
another notable contrast between the previous model and QGSJET II-03/04 is the steeper 
lateral distribution which gives the greater penetration in the Earth's atmosphere and 
higher value of shower maximum, $X_{\rm max}$, in the later versions. 

%\textcolor{red}{There were several shortcomings in the previous version of EPOS, 
%namely EPOS 1.99, which were resolved by utilizing the data from the Large Hadron 
%Collider (LHC) in the newer EPOS LHC version ~\cite{Pierog:2015EPOSatLHC}. However, 
%even in the updated EPOS LHC version (specifically the CORSIKA package version being 
%studied), discrepancies were found in the measured transverse momentum when comparing 
%it to the LHCb data for the anti-proton cross section in the large momentum range from 
%12 GeV/$c$ to 110 GeV/$c$ ~\cite{Graziani2017LHCb}. Using every high-energy model was 
%not possible for us as the simulation is computationally intensive. Given these 
%circumstances, we have decided not to work on  the EPOS model.}

We make six model combinations as follows:

\begin{enumerate}
\item {[A] GHEISHA \& QGSJET} 
\item {[B] UrQMD \& QGSJET} 
\item {[C] UrQMD \& QGSJET-II} 
\item {[D] GHEISHA \& SIBYLL} 
\item {[E] UrQMD \& SIBYLL} 
\item {[F] FLUKA \& SIBYLL} 
\end{enumerate}
We describe our studies in the following three subsections.

\subsection{Muon observables}
Here, we describe the results for the studies done on the muon
observables which are either measured or can be easily measured at ground
level detectors.
Same normalization factor is going for secondary fluxes which is obtained by
matching the input primary momentum distribution with the experimental data at
the top of atmosphere. Thus, simply the flux at ground is obtained in the same unit
as the measured flux at the top.
 The zenith angle and momentum spectra of muons obtained by CORSIKA are scaled 
by $1/\sin\theta$ to remove the effect of solid angle dependence on
zenith angle $\theta$.
Other than that we have only angle bin correction in which sampling is done
given by  the factor $F = (\pi/2)/\Delta\theta$, where $\Delta\theta$ is the
$\theta$ bin size. Here, $\pi/2$ is the total zenith angle coverage.   
  We calculate the $\chi^{2}/n$ (per degree of freedom) for each of the spectra given by 
\begin{eqnarray}
\chi^{2}/n = \frac{1}{n}\sum_{i=1}^{n} \Bigg(\frac{\phi_{i}^{\rm data} - \phi_{i}^{\rm corsika}}{\sqrt{(\Delta\phi_{i}^{\rm data})^2 + (\Delta\phi_{i}^{\rm corsika}})^2}\Bigg)^{2}.
\label{eq:chi}  
\end{eqnarray}
Here, $\phi_{i}^{\rm corsika} \pm \Delta\phi_{i}^{\rm corsika}$ is the value of the flux
with statistical error in CORSIKA and $\phi_{i}^{\rm data} \pm \Delta \phi_{i}^{\rm data}$
is the same in data for bin $i$ with total number of bins $n$.
Wherever experimental data was not available for a bin, the parametrization function was used
to interpolate between data points.

\begin{figure*}[h]
\begin{center}
\includegraphics[width=0.49\linewidth]{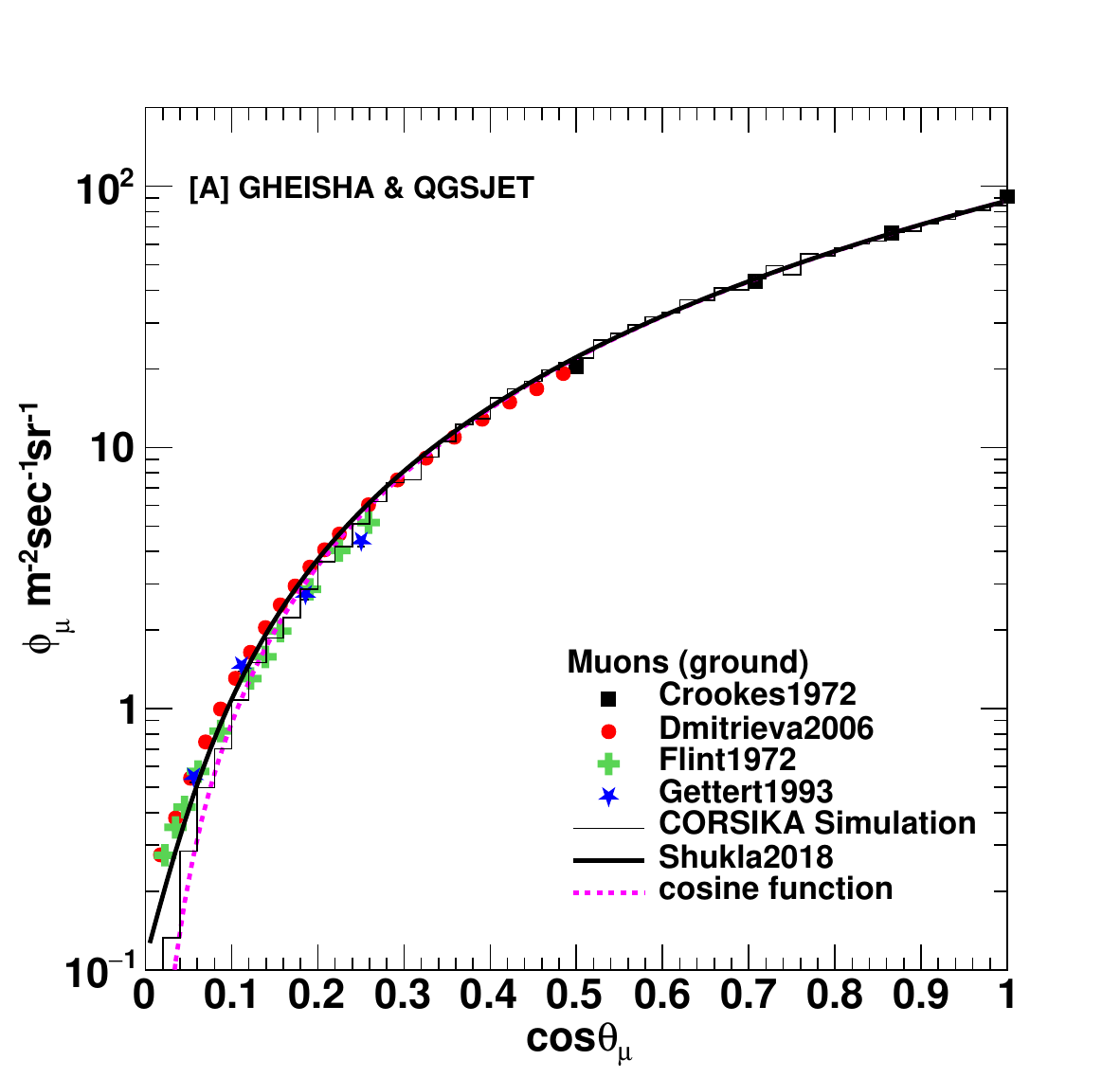}
\includegraphics[width=0.49\linewidth]{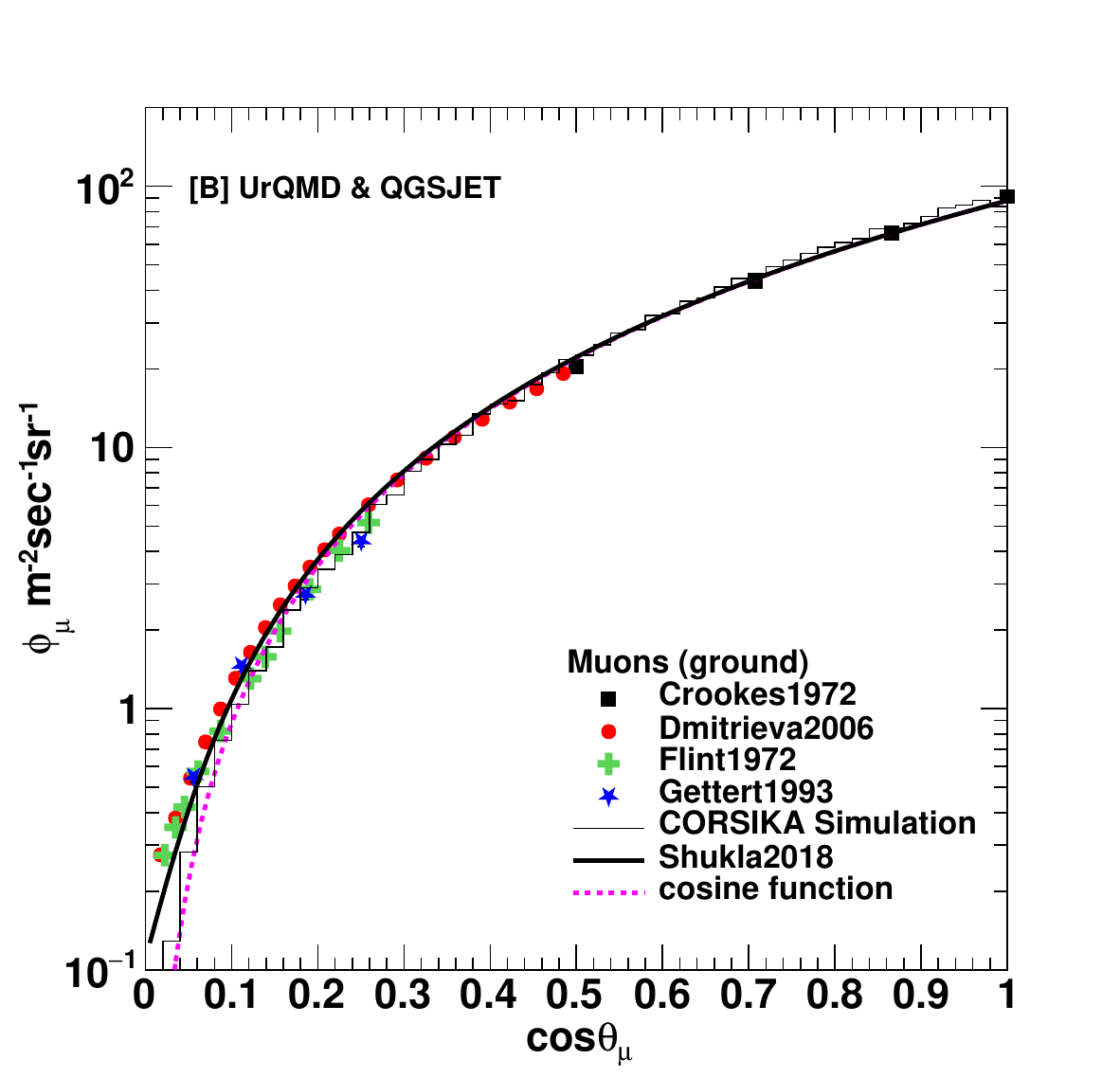} 
\includegraphics[width=0.49\linewidth]{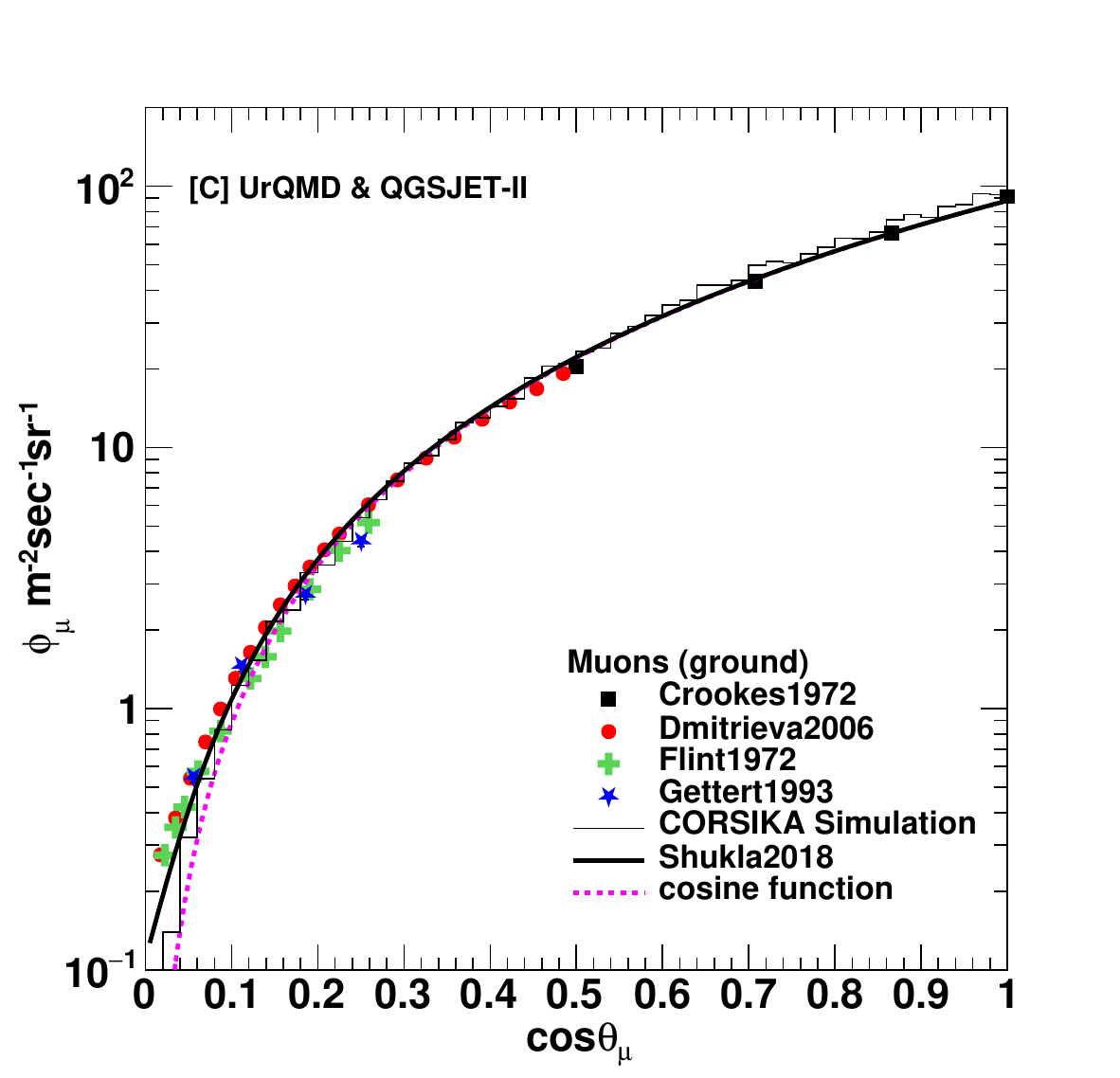}
\includegraphics[width=0.49\linewidth]{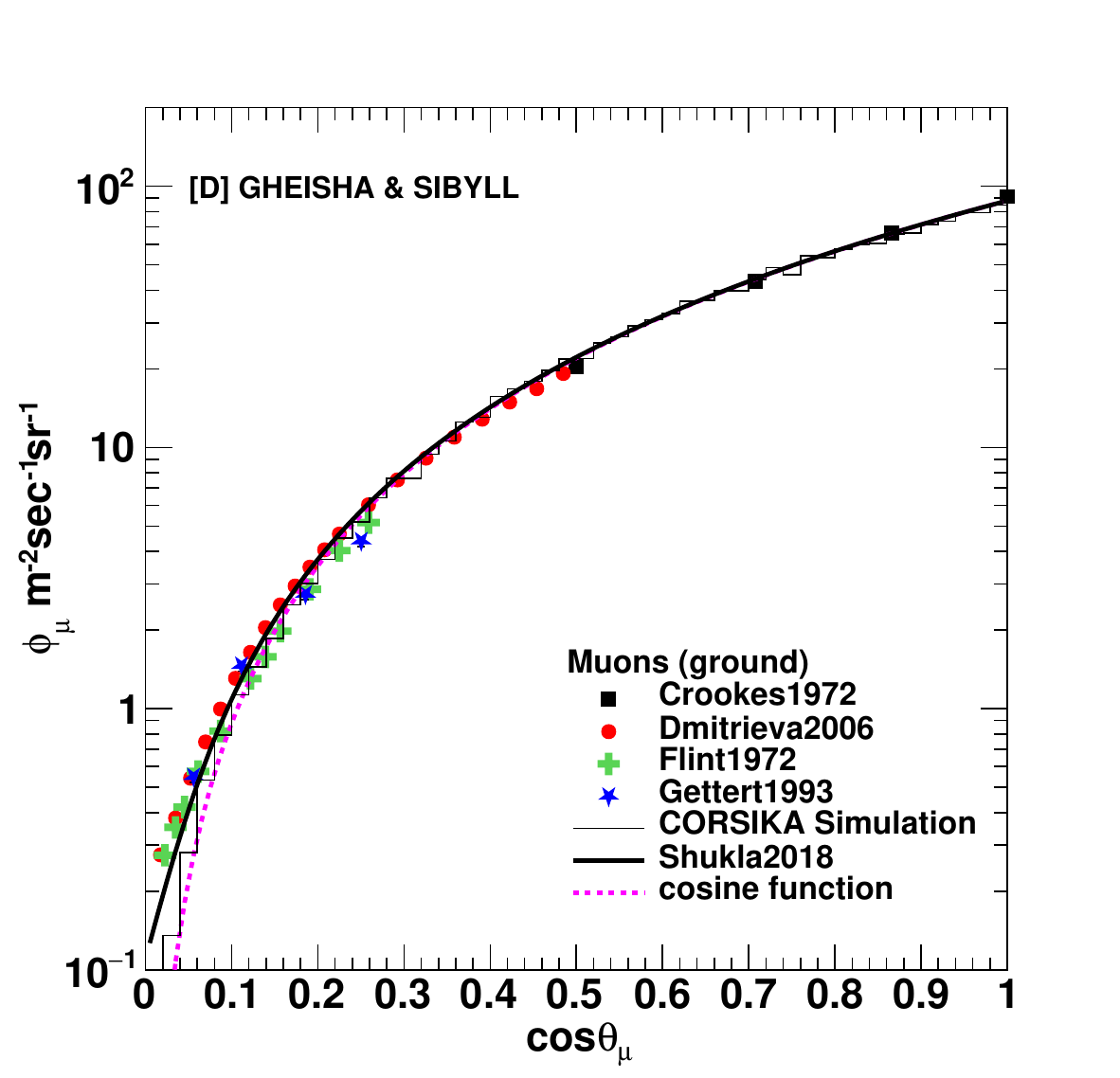}
\includegraphics[width=0.49\linewidth]{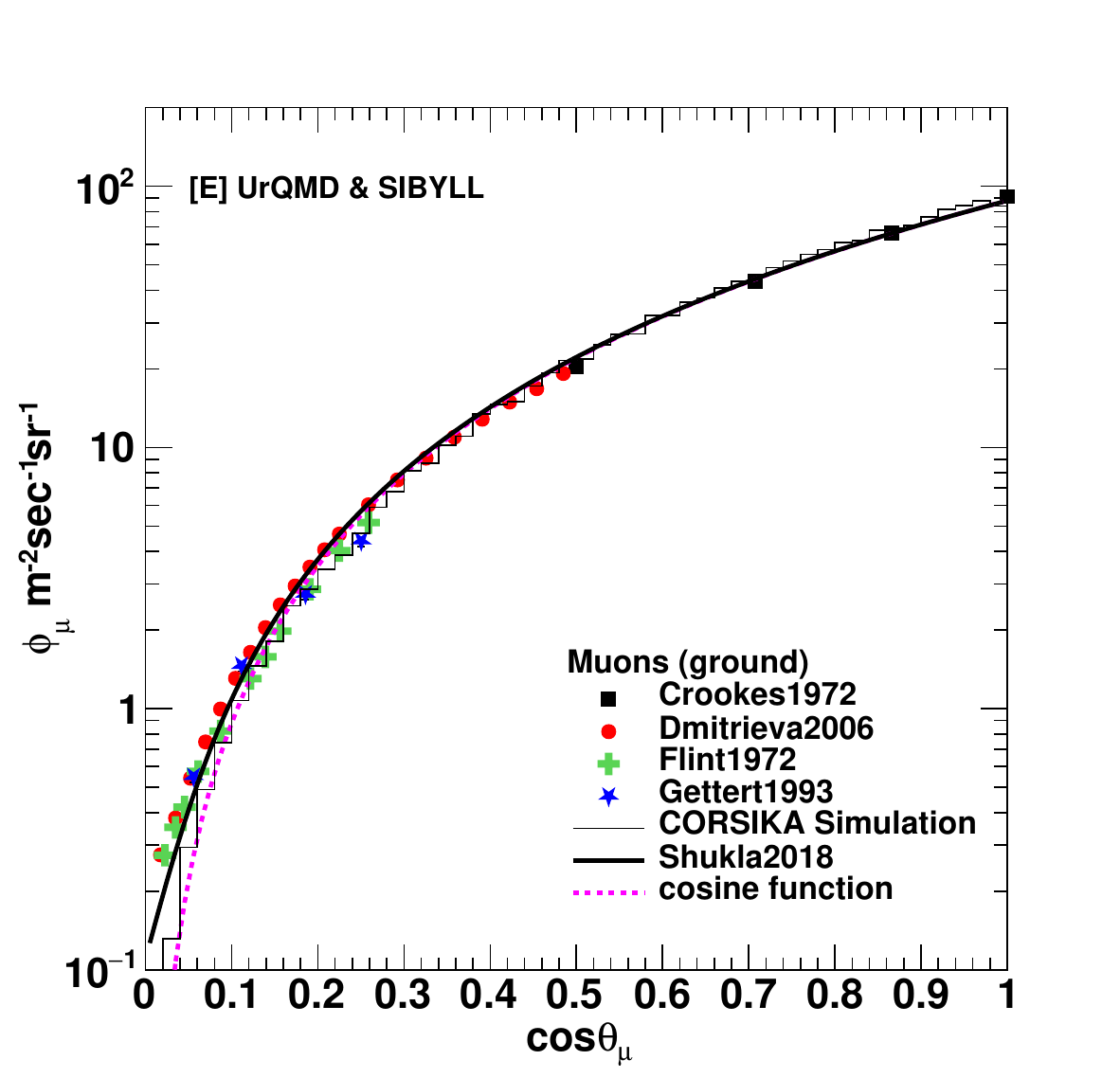}
\includegraphics[width=0.49\linewidth]{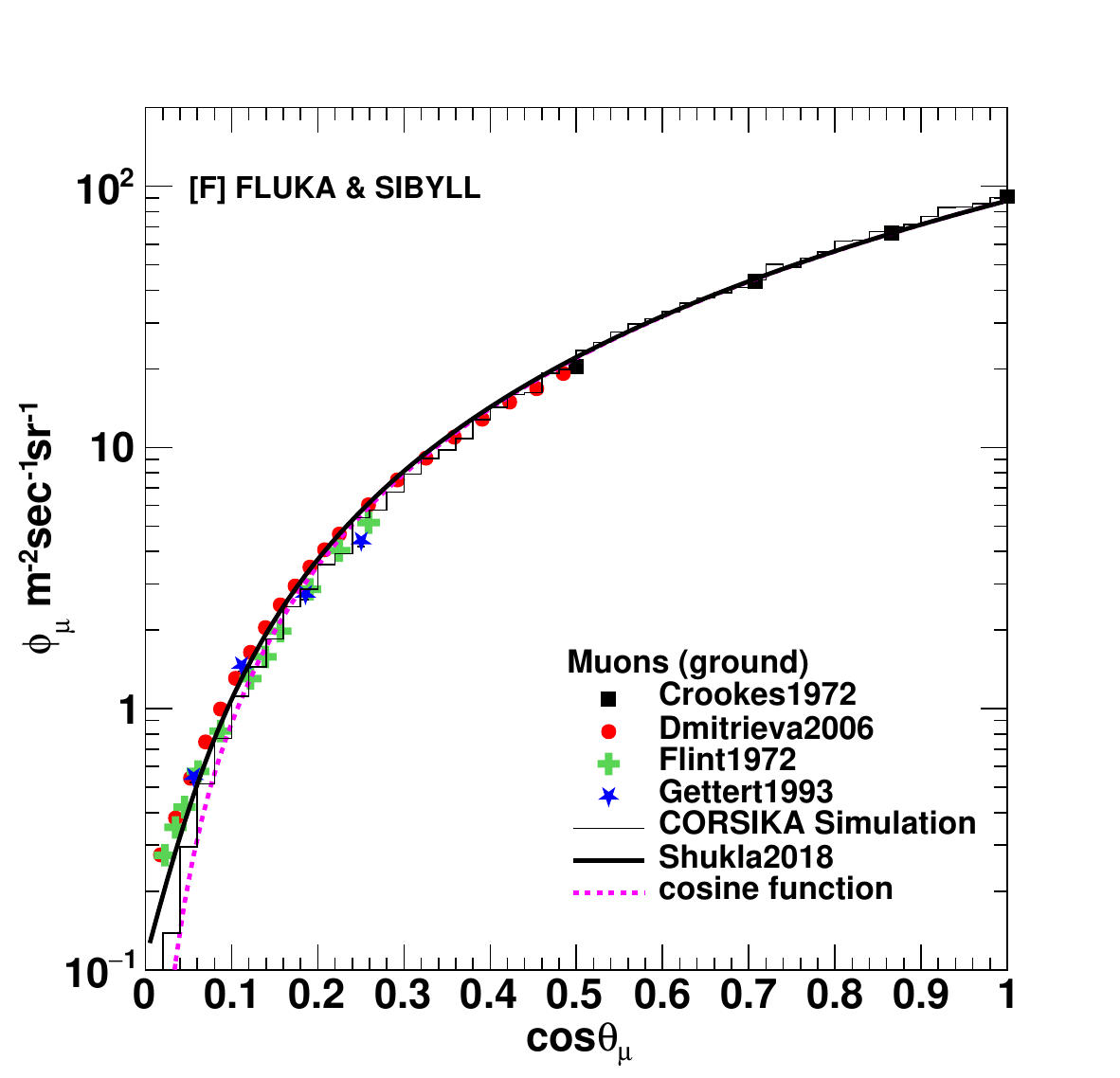}
\caption{The zenith angle distributions calculated from CORSIKA using 
various model combinations for muon momentum $p>0.5$ GeV/$c$ along with data measured 
by various experiments. The parameterization is given by Eq.~\ref{eq:ang} 
is also shown~\cite{shukla2016energy}. The data are taken from the collection of 
Ref.~\cite{cecchini2012atmospheric}.}
\label{figure2}	
\end{center}
\end{figure*}

Knowledge of muon zenith angle distributions, momentum distributions 
at zero and higher angles are important for the calibrations of the detectors at 
underground labs which are dedicated to measure the rare physics events. 
The shape and extent of these distributions depend on the material thickness. 
Thus they are useful in the estimations of depth of underground labs and 
detector material budget for such experiments.  
 Figure~\ref{figure2} shows the zenith angle distributions calculated with
CORSIKA using various model combinations for muon momentum $p$>0.5 GeV/$c$ along with
the data measured by various experiments. The parameterization given by Eq.~\ref{eq:ang} 
is also shown \cite{shukla2016energy}.
The data are taken from the collection of
Ref.~\cite{cecchini2012atmospheric} with the original references
~\cite{crookes1972investigation,dmitrieva2006measurements,flint1972variation,gettert1993}.
The different datasets have different muon energy thresholds and we take the
normalized data from the review~\cite{cecchini2012atmospheric}.
%The plots are scaled by $F = (\pi/{\rm 2})/\Delta x$, where $\Delta x$ = 0.02.
  Various models have differences in the hadronic multiplicity production 
and energy sharing among particles produced which actually affects the muon multiplicities 
at ground level and shower extent in the atmosphere. For example, in the QGSJET-II model, 
the shower depth $X_{\rm max}$ is larger than that in QGSJET and hence gives more 
contribution in the lower zenith angles.

\begin{figure*}[h]
\begin{center}
\includegraphics[width=0.49\linewidth]{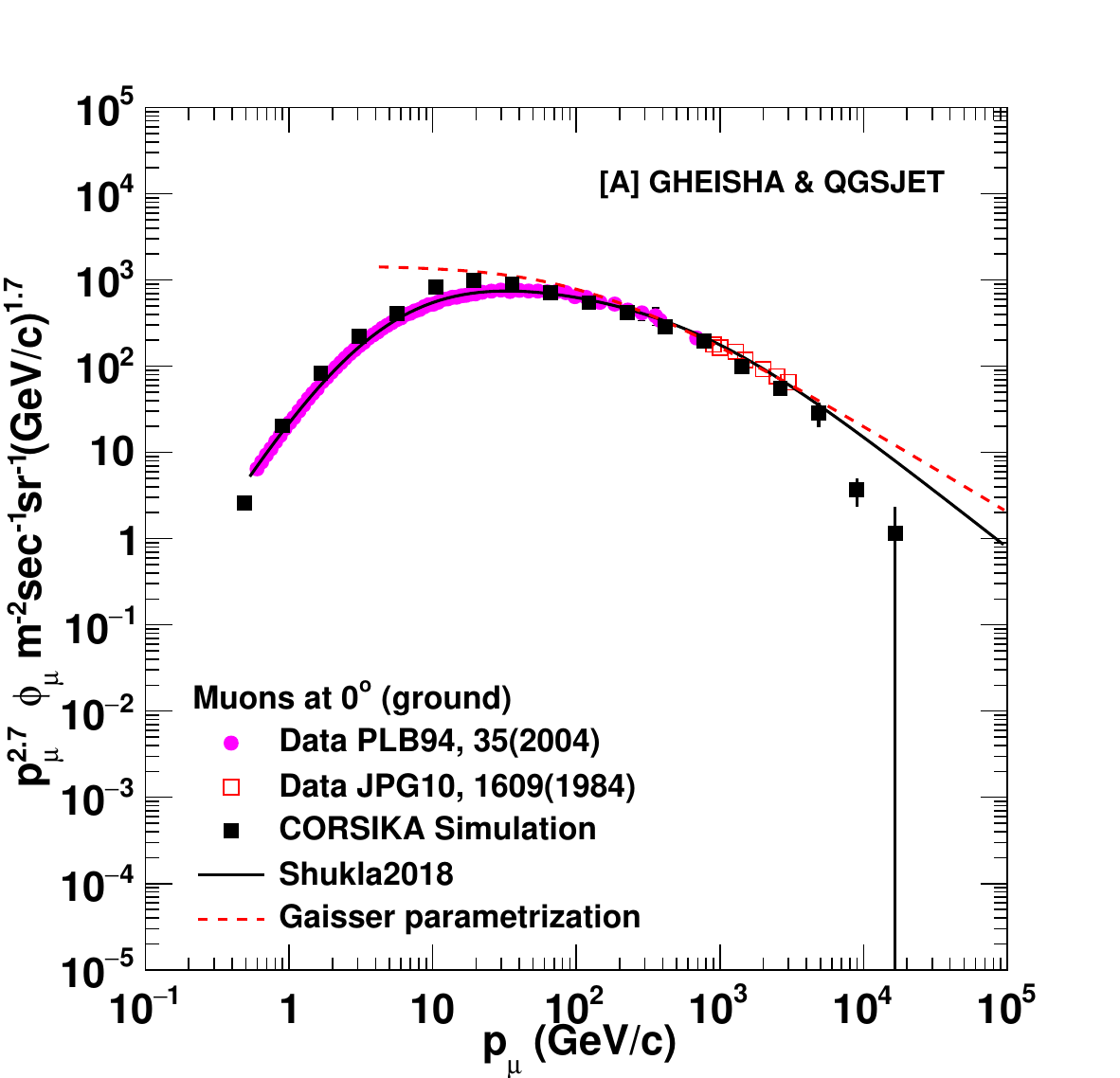}
\includegraphics[width=0.49\linewidth]{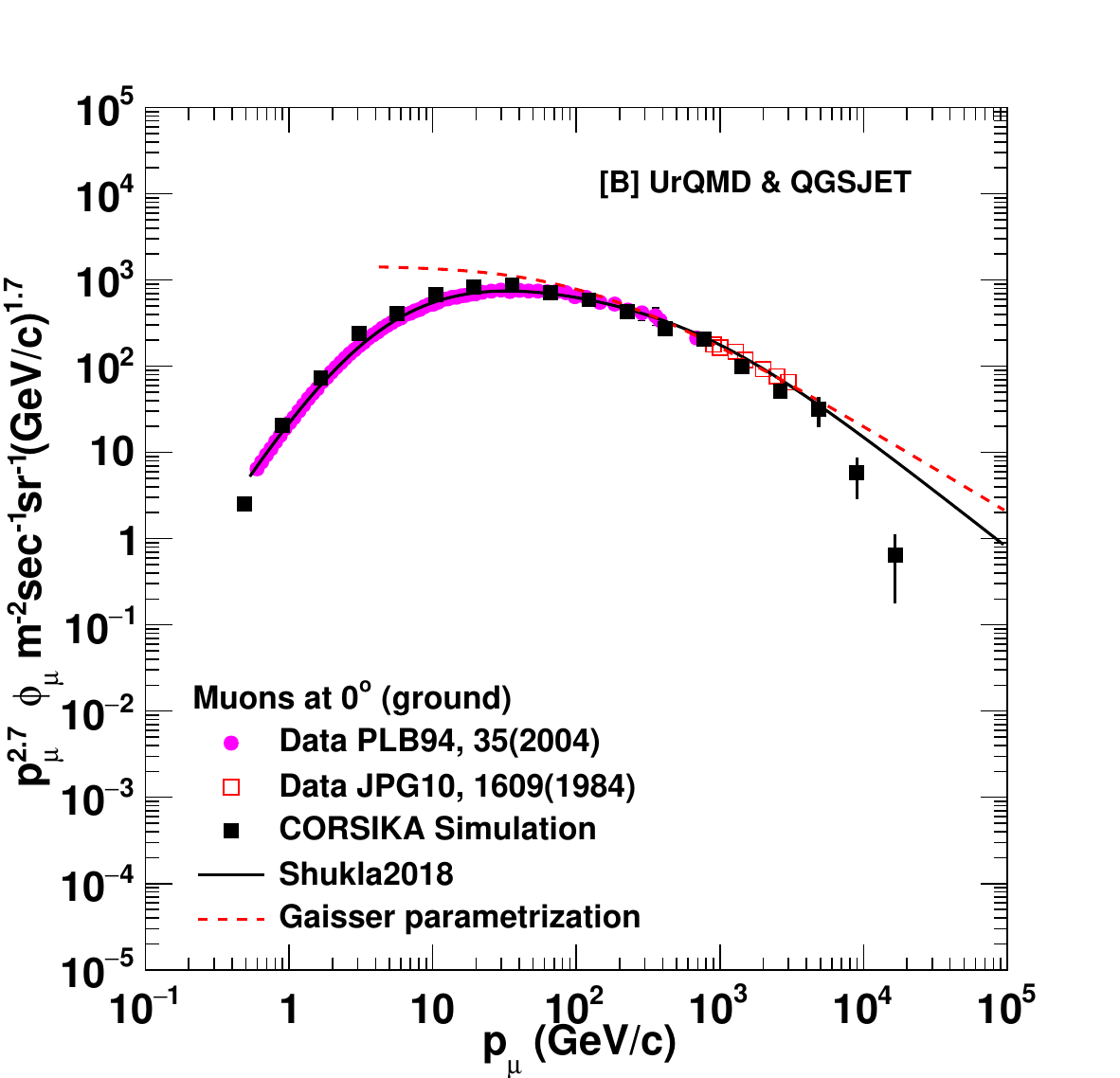} 
\includegraphics[width=0.49\linewidth]{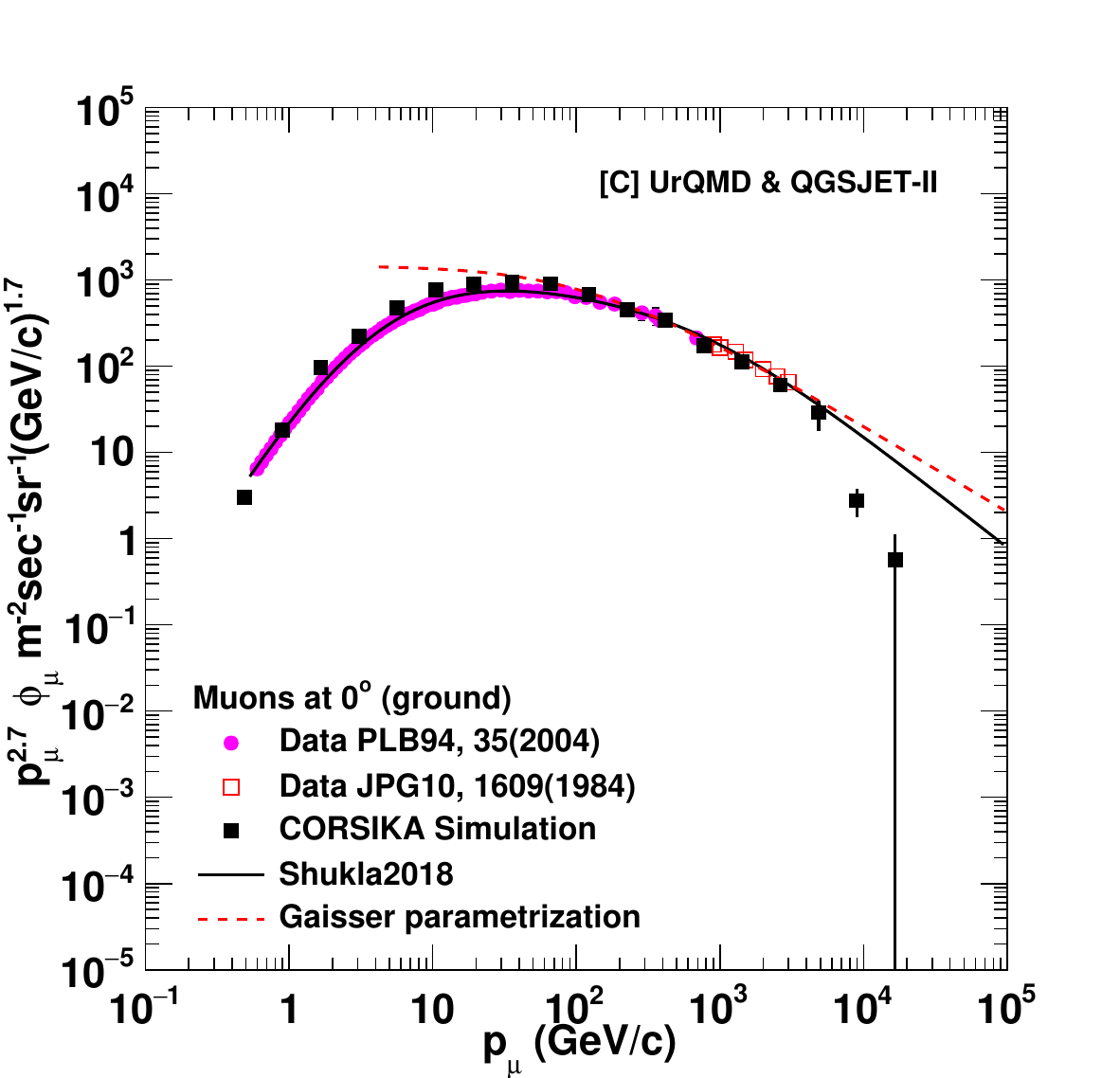}
\includegraphics[width=0.49\linewidth]{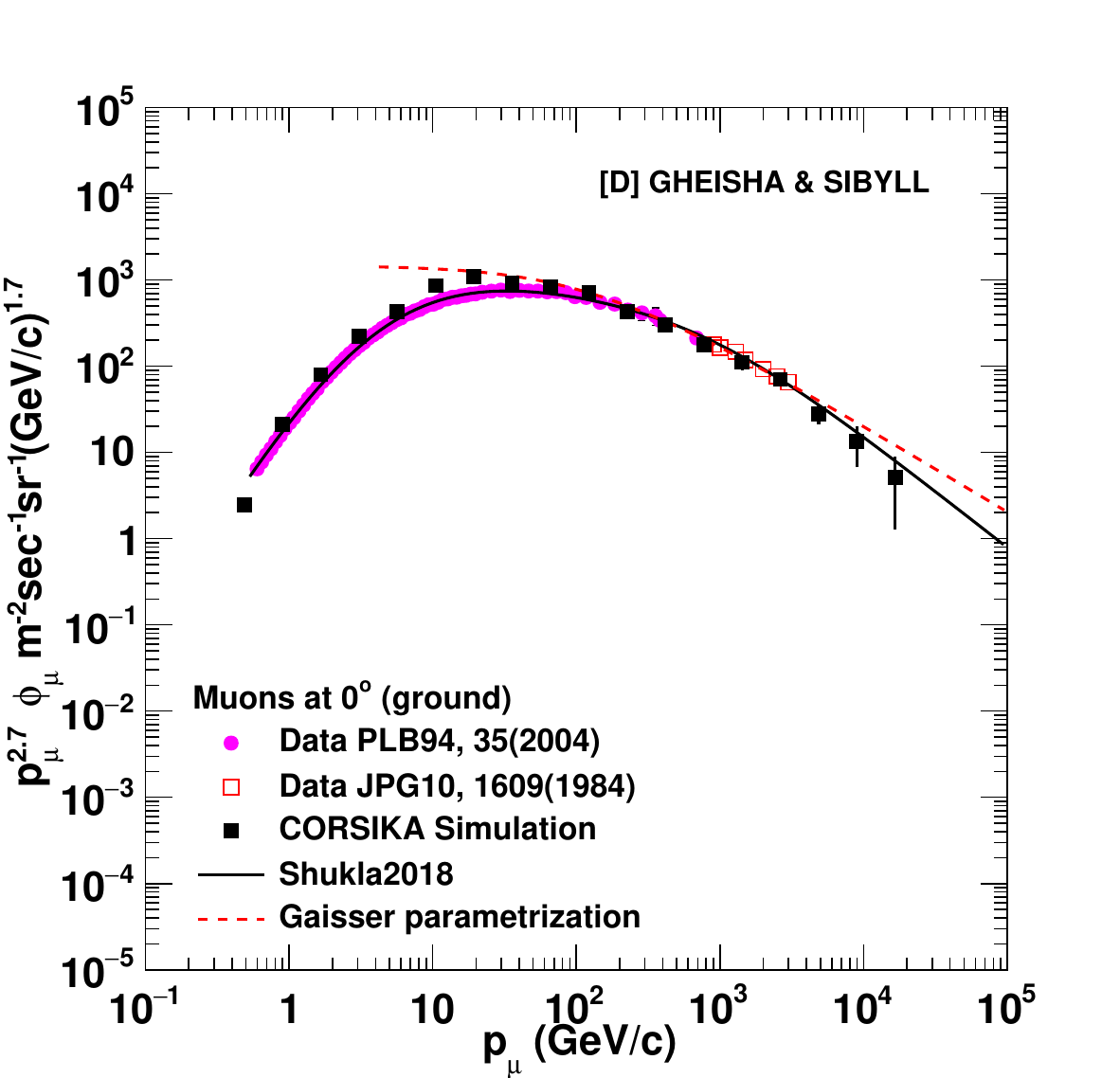}
\includegraphics[width=0.49\linewidth]{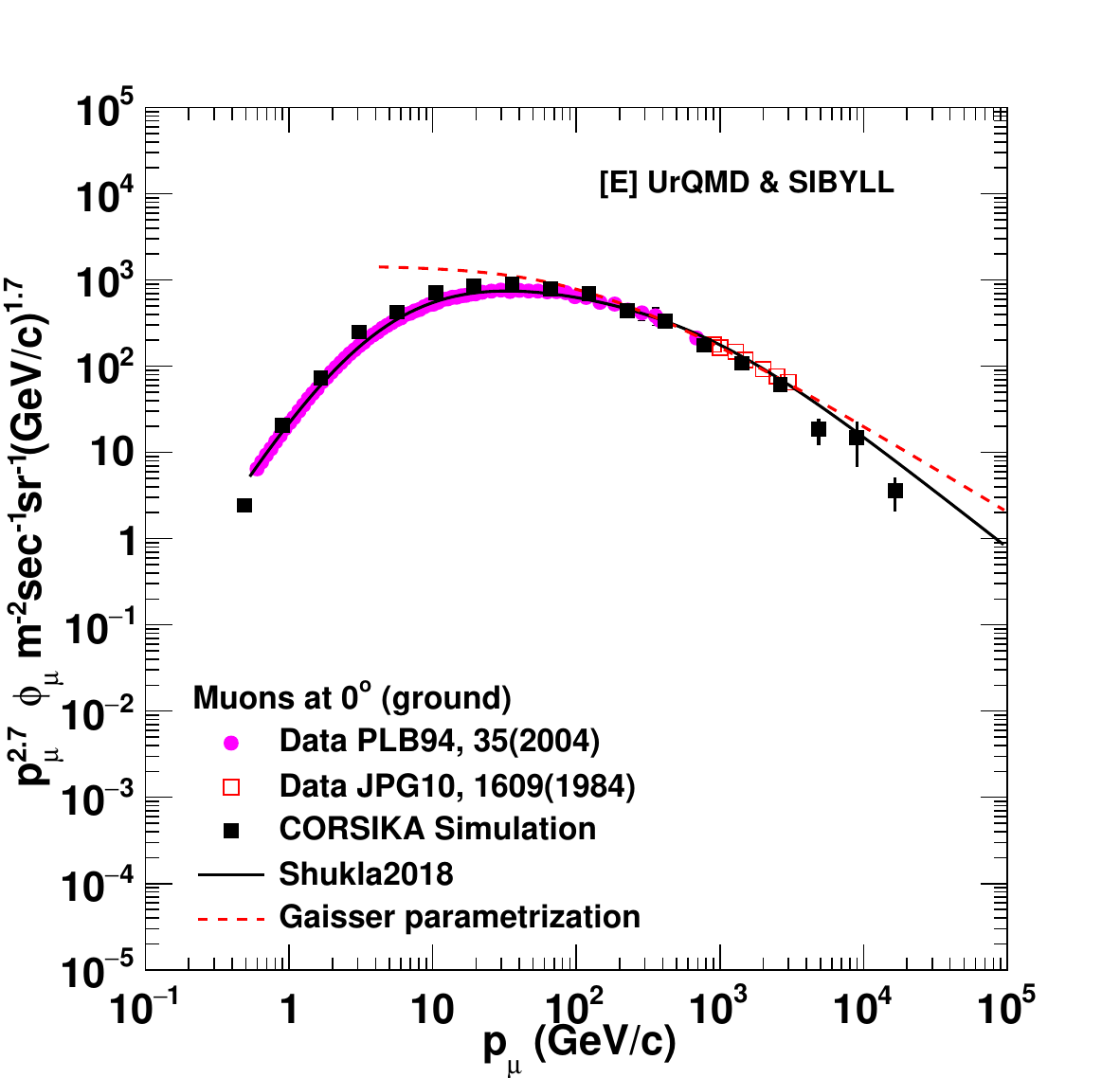}
\includegraphics[width=0.49\linewidth]{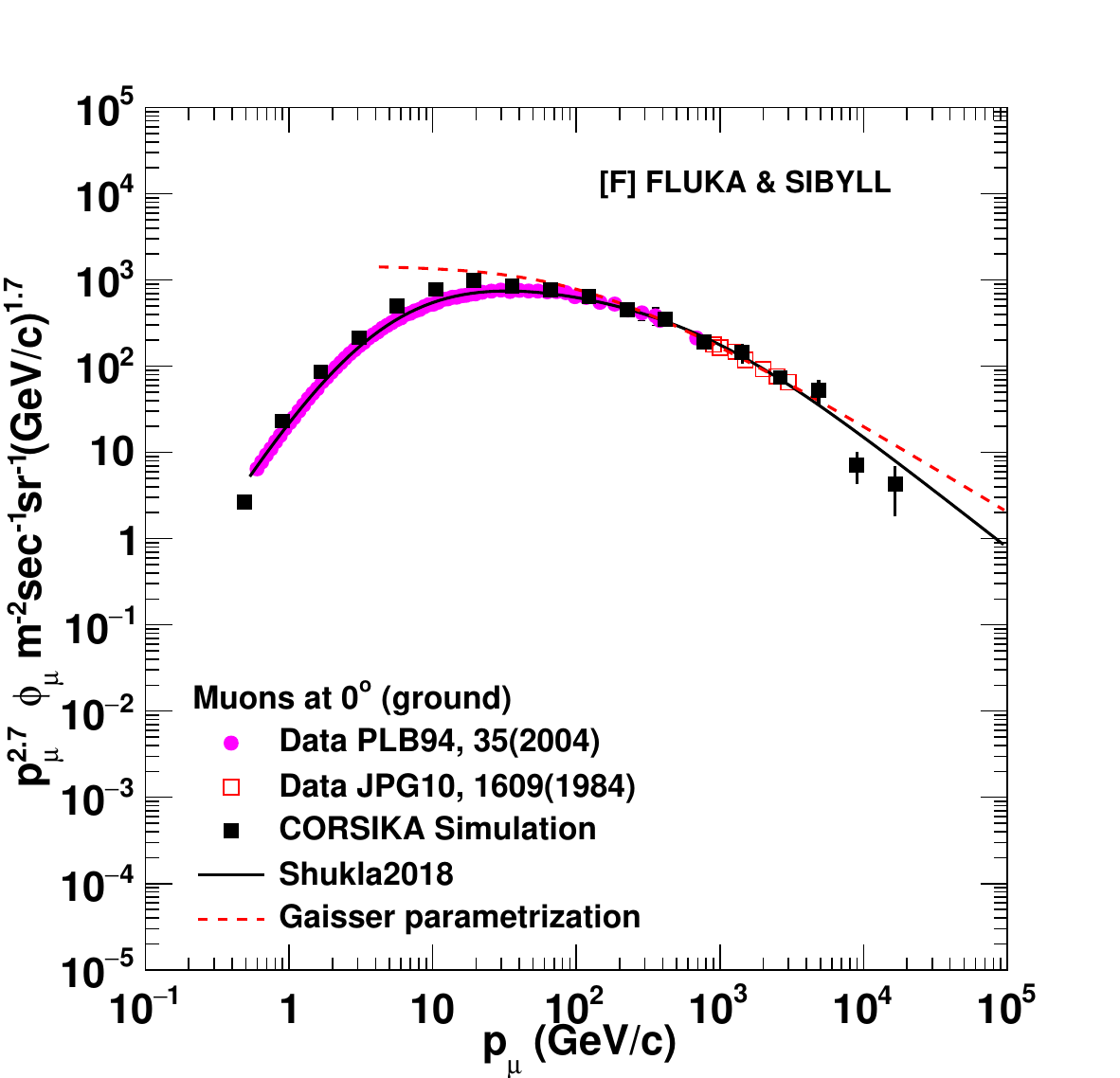}
\caption{The momentum distribution of vertical atmospheric muons 
at the ground level calculated by CORSIKA using different 
hadronic interaction model combinations for muon momentum $p>0.5$ GeV/$c$ along with the 
experimental data~\cite{haino2004measurements,rastin1984accurate}. 
The parametrizations given by Eq.\ref{eq:shukla} \cite{shukla2016energy}
and Eq.\ref{eq:gaisser} \cite{gaisser2002semi} are also shown.}
\label{figure3}
\end{center}
\end{figure*}

\begin{figure*}[h]
\begin{center}
\includegraphics[width=0.49\linewidth]{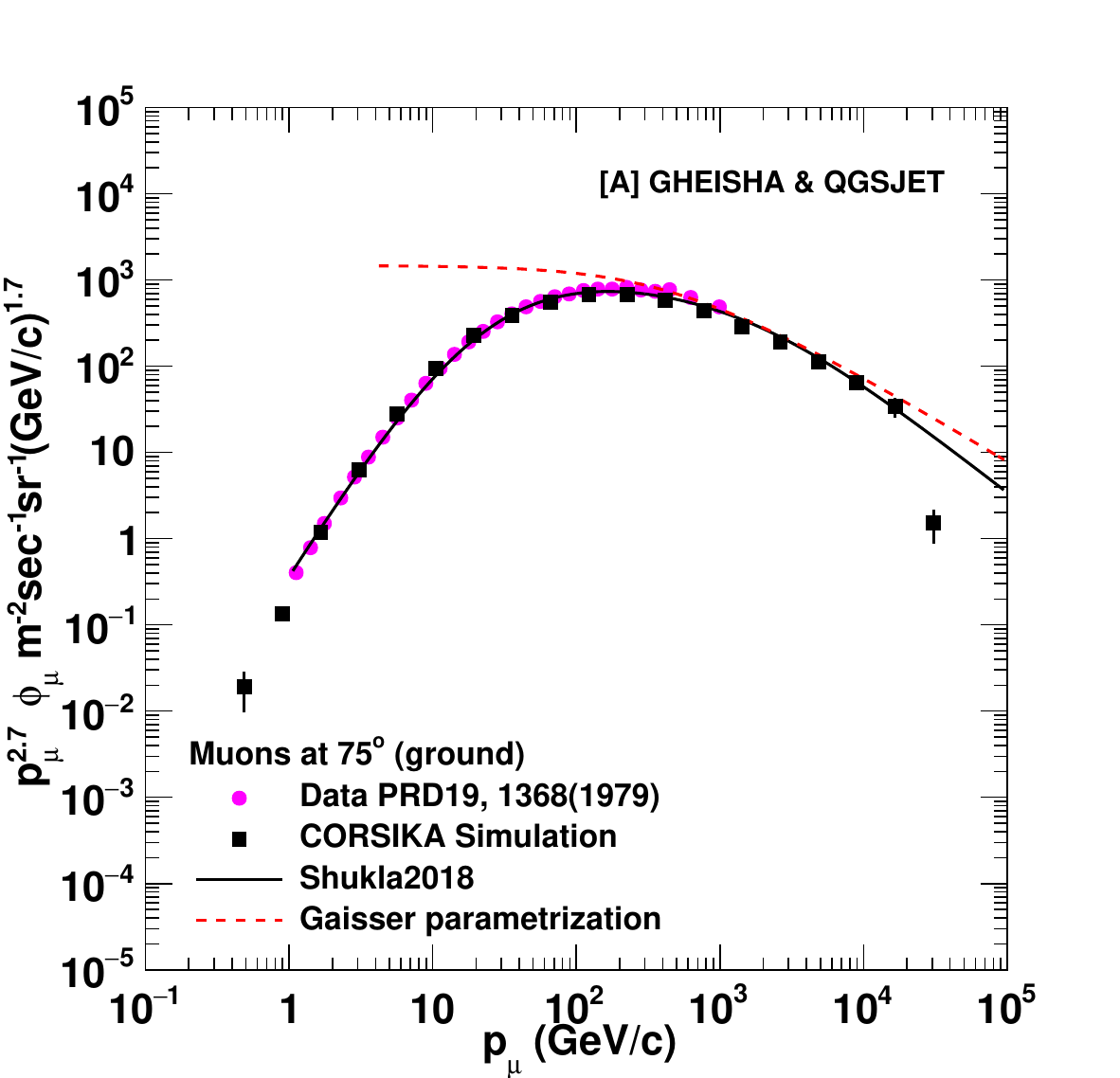}
\includegraphics[width=0.49\linewidth]{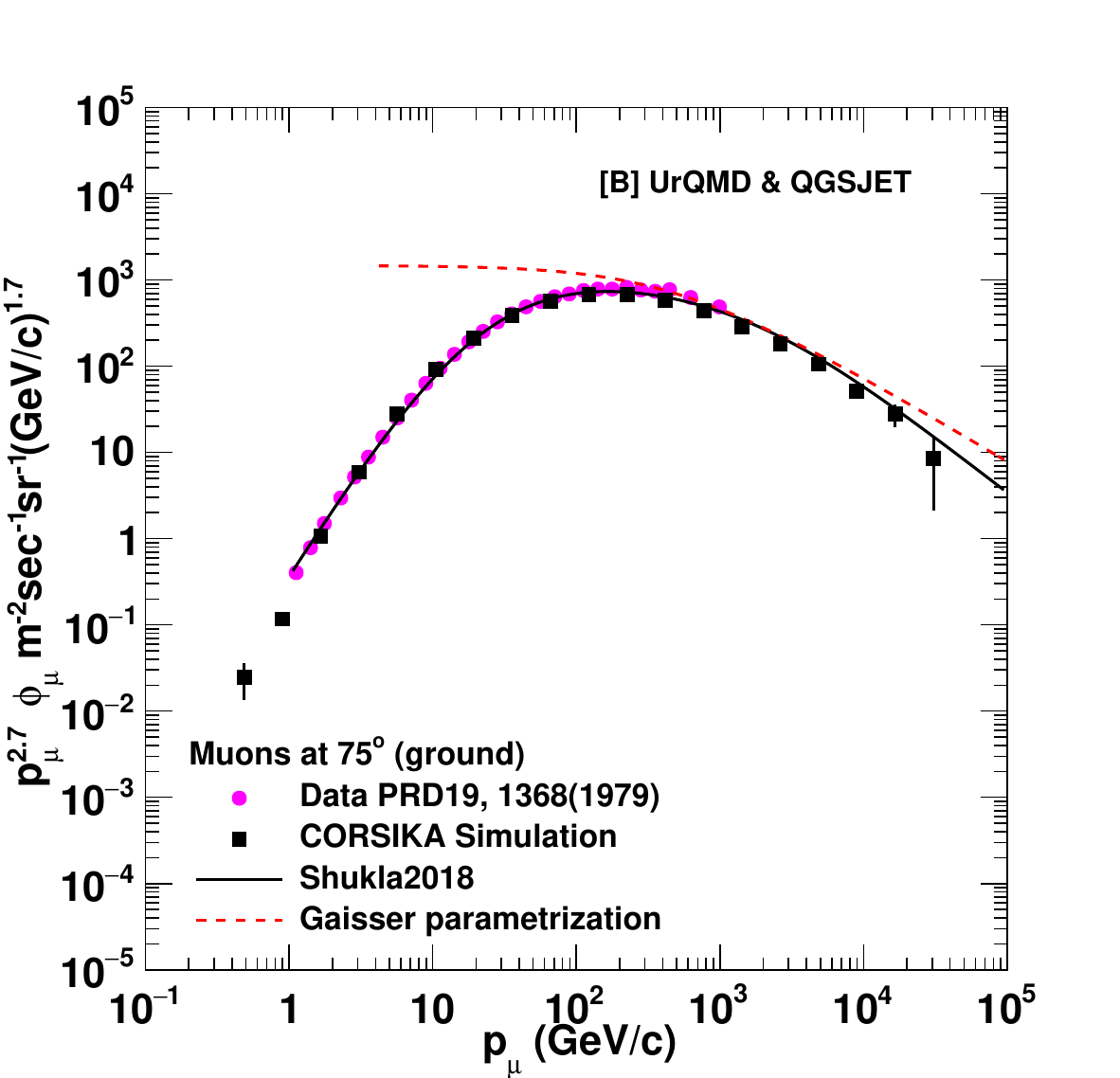}
\includegraphics[width=0.49\linewidth]{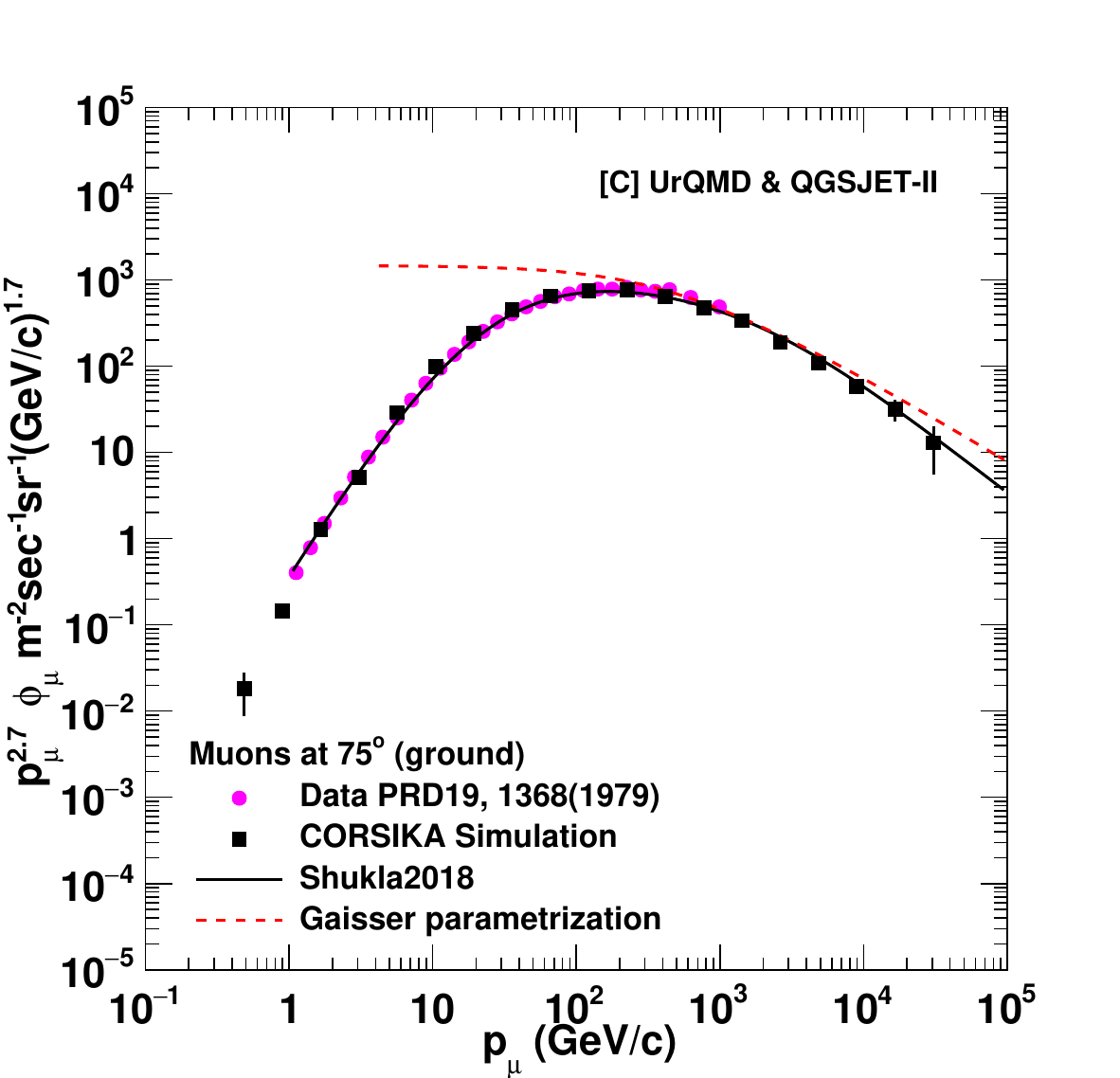}
\includegraphics[width=0.49\linewidth]{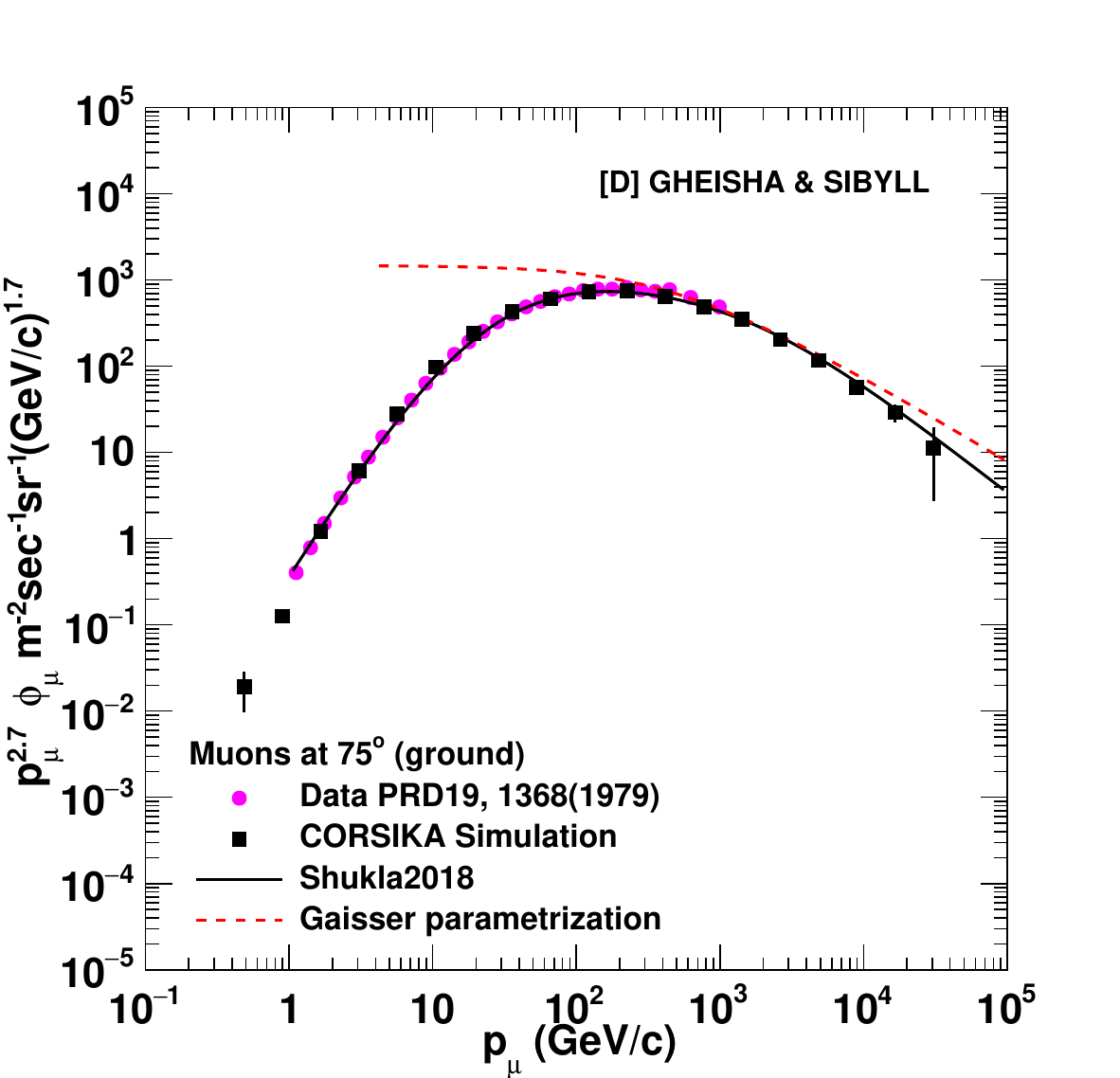}
\includegraphics[width=0.49\linewidth]{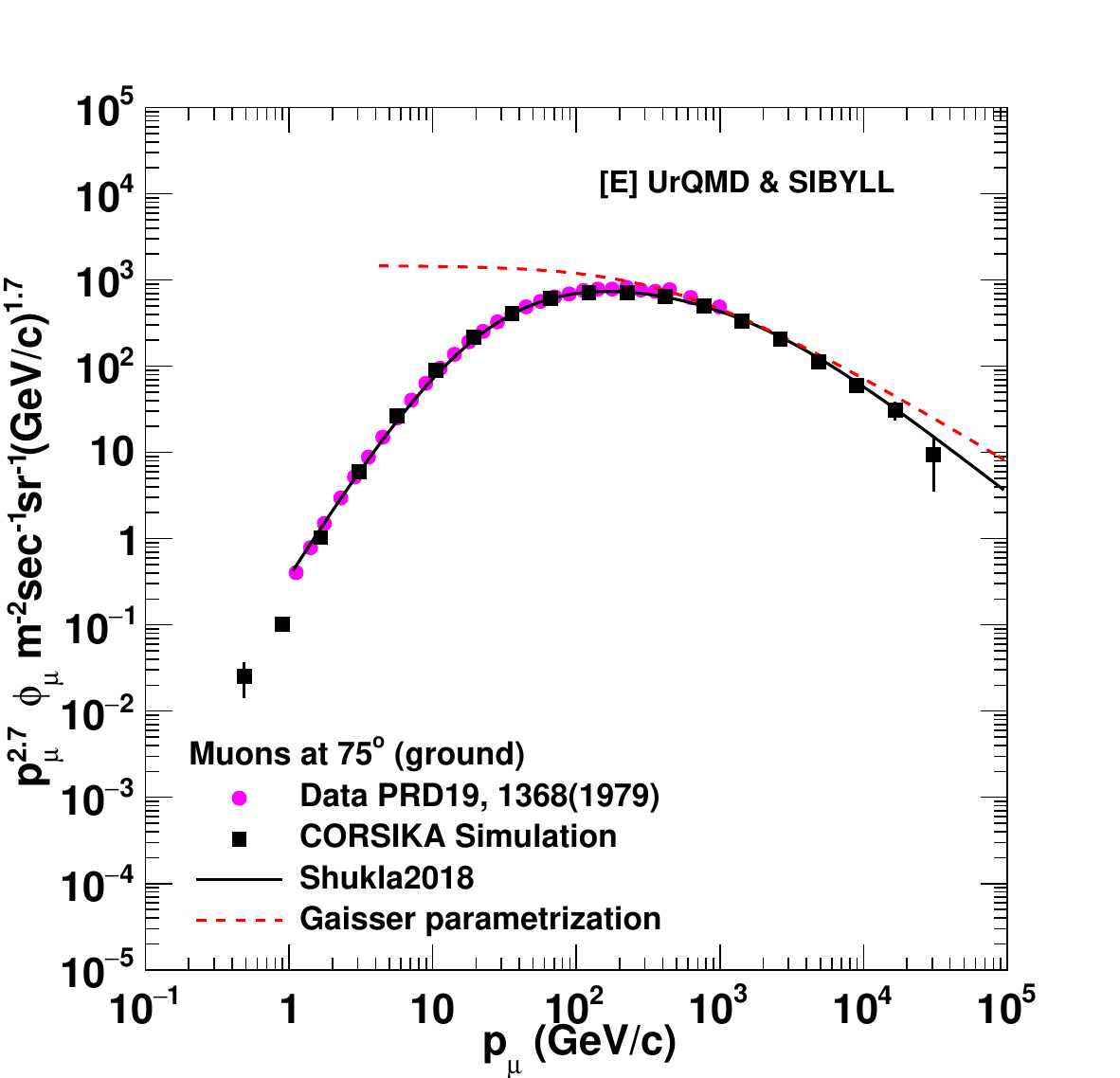}
\includegraphics[width=0.49\linewidth]{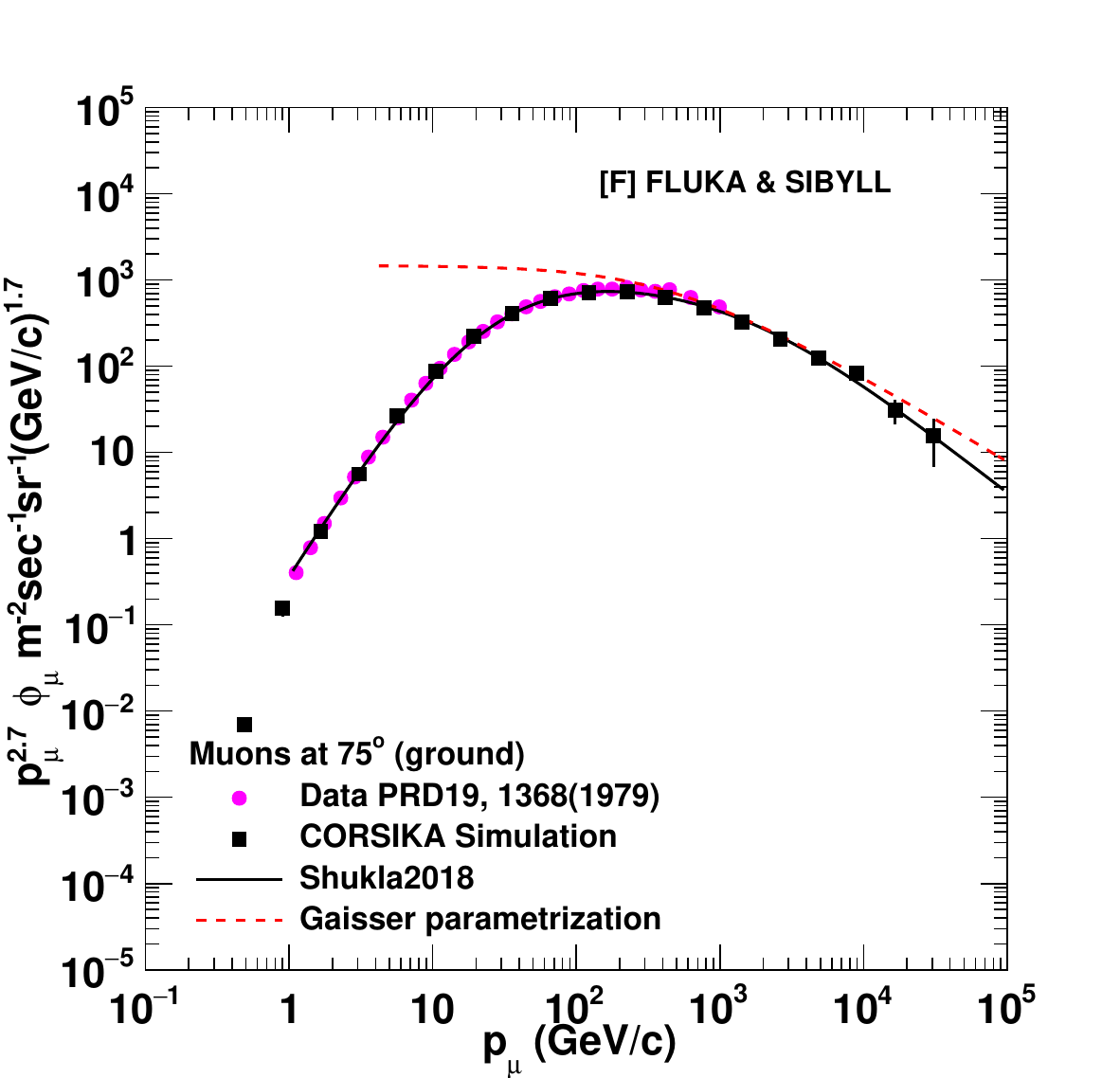}
\caption{The momentum distribution of atmospheric muons at $75^{\circ}$ 
zenith angle at the ground level calculated by CORSIKA using different 
hadronic interaction model combinations for $p>0.5$ GeV/$c$ along with 
the experimental data~\cite{jokisch1979cosmic}. The parametrizations 
given by Eq.~\ref{eq:shukla} and Eq.~\ref{eq:gaisser} are also shown.}
\label{figure4}	
\end{center}
\end{figure*}

Figure~\ref{figure3} shows the momentum distribution of vertical atmospheric muons 
at the ground level calculated by CORSIKA using 
different hadronic interaction model combinations for $p>0.5$ GeV/$c$ along 
with the experimental data~\cite{haino2004measurements,rastin1984accurate}.
The parametrizations given by Eq.\ref{eq:shukla} and Eq.\ref{eq:gaisser} are 
also shown.
The plots in Figure \ref{figure3} are obtained by integrating all muons
falling between 0 to 10 degrees and are scaled by
$F = (\pi/2)/\Delta\theta$, where $\Delta\theta=10^\circ$.
All model combinations give a good description of experimental
data although all underestimate the muon flux in the very low energy region.
The parametrization by Eq.\ref{eq:shukla} gives a good description of the data.

Figure~\ref{figure4} shows the momentum distribution of atmospheric muons 
at $75^{\circ}$ zenith angle at the ground level calculated by CORSIKA 
using the different hadronic interaction model combinations for $p>0.5$ GeV/$c$ 
along with the experimental data~\cite{jokisch1979cosmic}. The 
parametrizations given by Eq.~\ref{eq:shukla} and Eq.~\ref{eq:gaisser}
are also shown.
The plots in Figure \ref{figure4} are obtained by integrating all muons
falling between 71 to 81.8 degrees and are scaled by 
$F = (\pi/2)/\Delta\theta$, where $\Delta\theta=10.8^\circ$.
All model combinations give a good description of experimental
data although all underestimate the muon flux in the very low energy region.
Models [D], [E] and [F] give a good description of the data.
The parametrization by Eq.~\ref{eq:shukla} gives a good description of the data.

The $\chi^{2}/n$ calculated (with Eq.~\ref{eq:chi}) for all the models
corresponding to Figures \ref{figure2}, \ref{figure3} and \ref{figure4}
are given in Table~\ref{table5}. 
 For $\chi^{2}/n$ calculations the muon cosine zenith angle range from 0.1 to 1
and the momentum range from 1 GeV/$c$ to 5 TeV/$c$ are included. 
Wherever experimental data was not available, the parametrization function was used 
to interpolate between data points.
 All the model combinations produce the shape of zenith angle distribution
given by $\cos^{\rm 2}\theta$ and models [A], [C] and [D] give good descriptions of the data.
The mismatch at higher angles is due to flat Earth approximation used in CORSIKA and
this mismatch exists for all model combinations.
The parametrization given by Eq.~\ref{eq:ang} produces the shape of the distribution 
at all angles.

%From this table it is noted that all models other than model combination [C]
%give good values of $\chi^{2}$. The model QGSJET-II is an improved
%version of the QGSJET and might be suitable for ultra-high energies. 
%Overall, the model combinations
%[E] and [F] give the best results as compared to the data. 

\begin{table*}[h]
\begin{center}
\caption{Calculated $\chi^{\rm 2}/n$ for the cosine of zenith angle distribution and momentum 
 distributions at 0$^{\circ}$ and 75$^{\circ}$ from CORSIKA models.}
\label{table5}
\resizebox{\columnwidth}{!}{%
\begin{tabular}{l c c c} 
\hline
\textbf{Model} & \multicolumn{3}{c}{\textbf{$\chi^{\rm 2}/n$}}  \\
\cline{2-4}
\textbf{Combination} & \textbf{for $\theta$ distribution} & \textbf{for momentum at 0$^{\circ}$} & \textbf{for momentum at 75$^{\circ}$} \\
\hline
\textbf{[A] GHEISHA \& QGSJET} & 4.28 & 1.62 & 8.83 \\
\textbf{[B] UrQMD \& QGSJET}   & 6.52 & 1.23 & 7.46 \\
\textbf{[C] UrQMD \& QGSJET-II}& 4.24 & 2.39 & 6.63 \\
\textbf{[D] GHEISHA \& SIBYLL} & 2.79 & 1.80 & 3.52 \\
\textbf{[E] UrQMD \& SIBYLL}   & 5.69 & 1.69 & 1.09 \\
\textbf{[F] FLUKA \& SIBYLL}   & 5.20 & 1.43 & 1.47 \\
\hline
\end{tabular}
}
\end{center}
\end{table*}

The integrated muon fluxes for all the model combinations calculated from 
zenith angle distribution and from momentum distributions at $0^{\circ}$ 
and $75^{\circ}$ are given in the Table~\ref{table4}.
The integrals obtained from theta distribution are multiplied with
the $2\pi.(\pi/2)$ to cover the whole solid angle.
We find that the integrated muon flux given by all interaction models
reasonably match with each other except for combination [C] UrQMD \& QGSJET-II
which is higher in all 3 cases. 
The effect of deeper shower maximum in case of QGSJET-II with 
respect to QGSJET can be easily seen in the muon zenith angle and momentum 
distributions and in the integrated values in the Table~\ref{table4} [B] and [C] models.

\begin{table*}[h]
\begin{center}
\caption{Integrated muon flux for different models at ground level for muon momentum $p>0.5$ GeV/$c$.}    
\label{table4}
\resizebox{\columnwidth}{!}{%
\begin{tabular}{l c c c} 
\hline
& \multicolumn{3}{c}{\textbf{Integrated muon flux}}  \\
\cline{2-4}
\textbf{Model} & \textbf{for $\theta$ distribution} & \textbf{for momentum at $0^{\circ}$} & \textbf{for momentum at $75^{\circ}$} \\
\textbf{Combination} & (m$^{-2}$sec$^{-1}$)  &  (m$^{-2}$sec$^{-1}$sr$^{-1}$)  &  (m$^{-2}$sec$^{-1}$sr$^{-1}$) \\  
\hline
\textbf{[A] GHEISHA \& QGSJET} & 184.80 & 75.55 & 1.32\\
\textbf{[B] UrQMD \& QGSJET}   & 191.49 & 73.20 & 1.28\\
\textbf{[C] UrQMD \& QGSJET-II}& 197.59 & 80.38 & 1.35\\
\textbf{[D] GHEISHA \& SIBYLL} & 183.29 & 74.97 & 1.33\\
\textbf{[E] UrQMD \& SIBYLL}   & 190.00 & 73.19 & 1.25\\
\textbf{[F] FLUKA \& SIBYLL}   & 189.85 & 80.74 & 1.26\\
\hline
\end{tabular}
}
\end{center}
\end{table*}

\begin{figure*}[h]
\begin{center}
\includegraphics[width=0.49\linewidth]{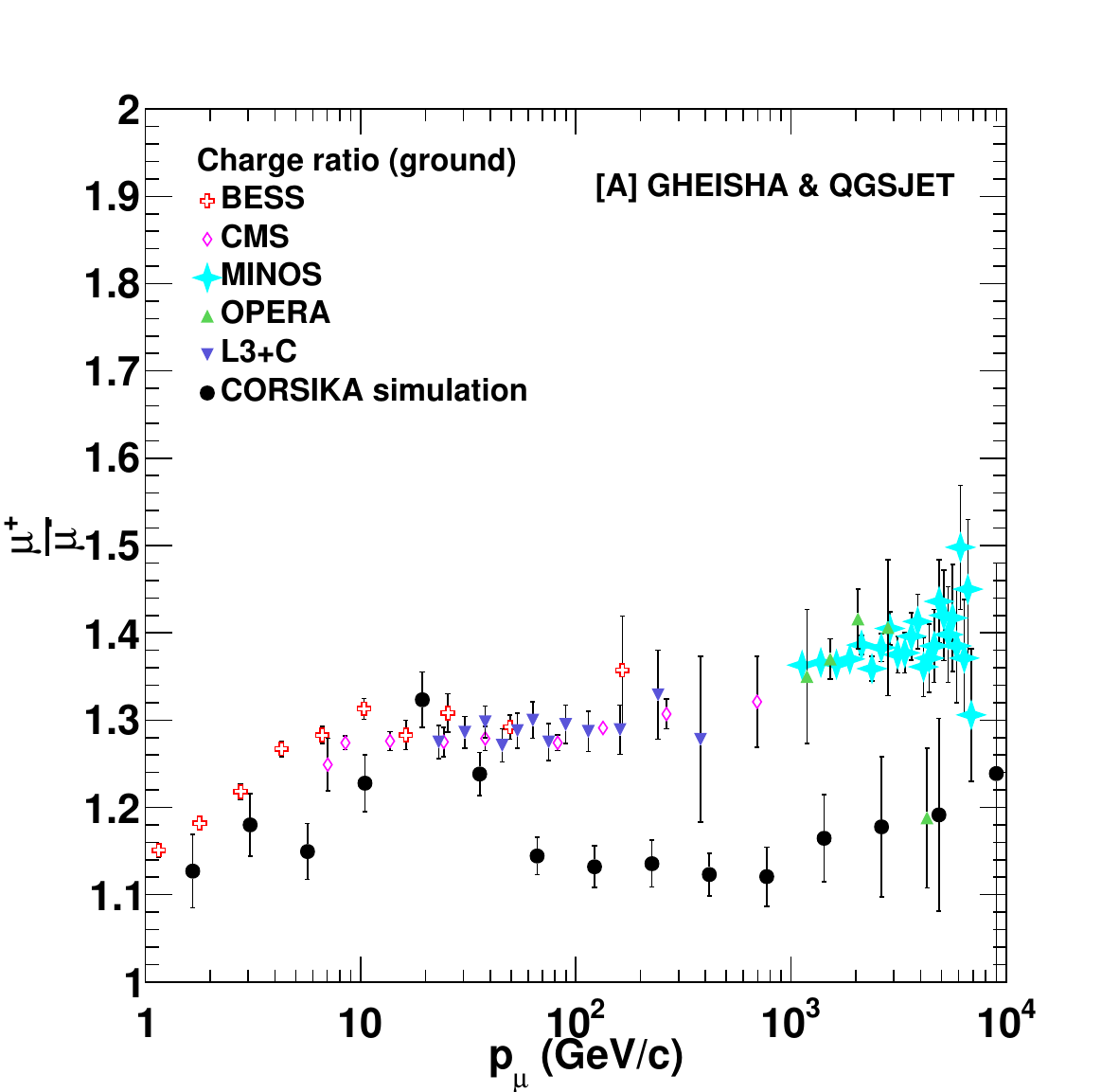}
\includegraphics[width=0.49\linewidth]{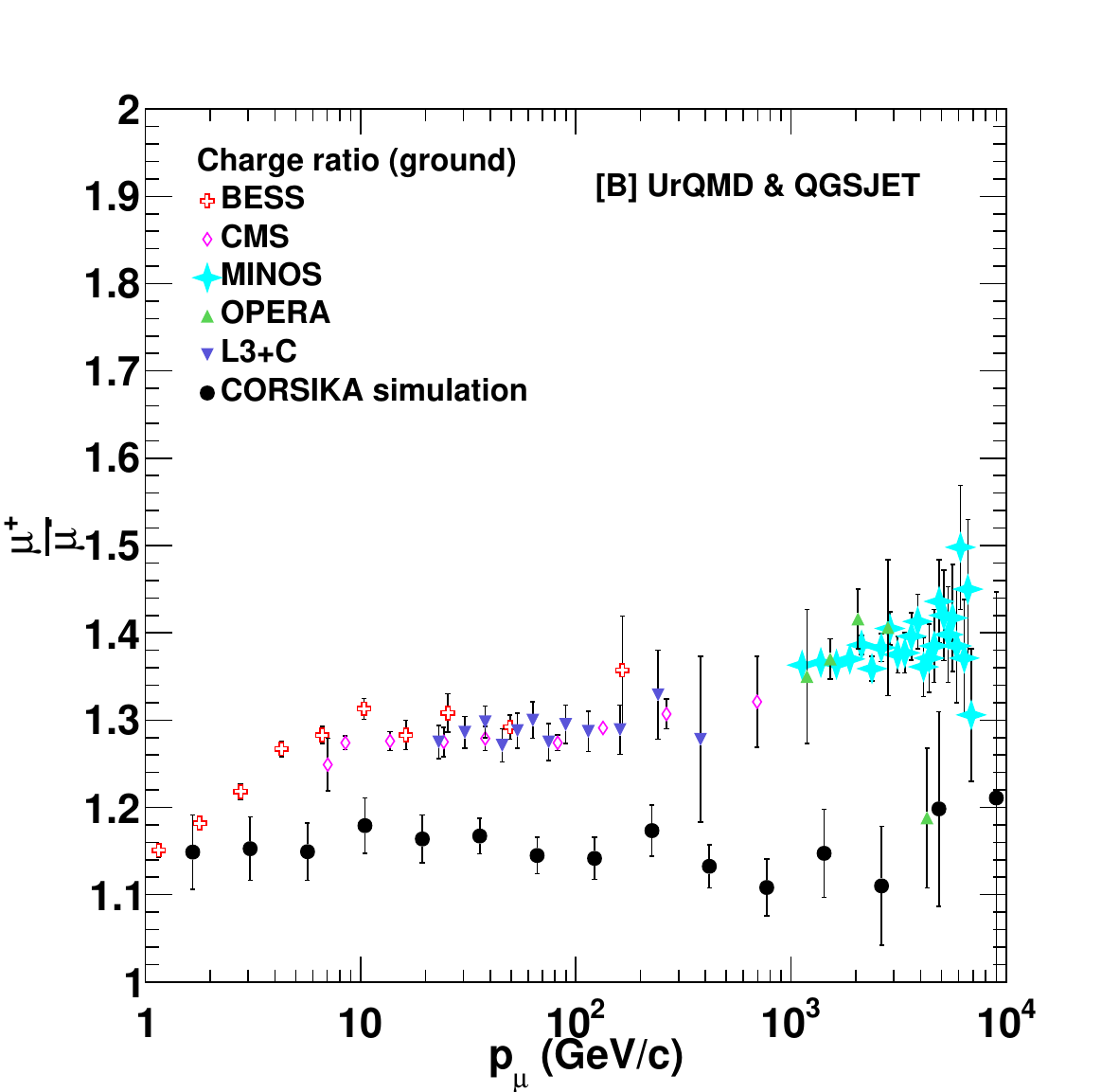}
\includegraphics[width=0.49\linewidth]{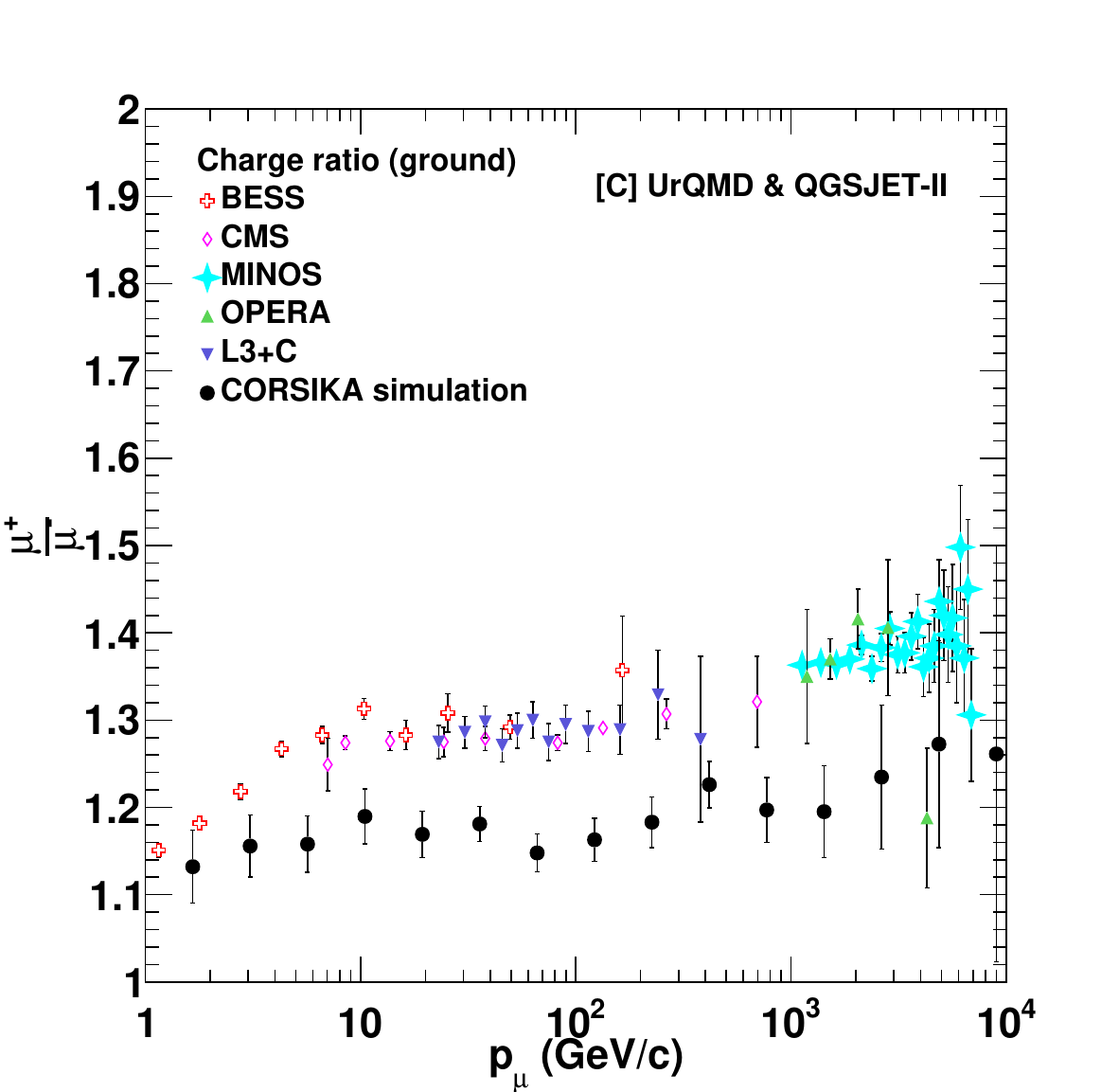}
\includegraphics[width=0.49\linewidth]{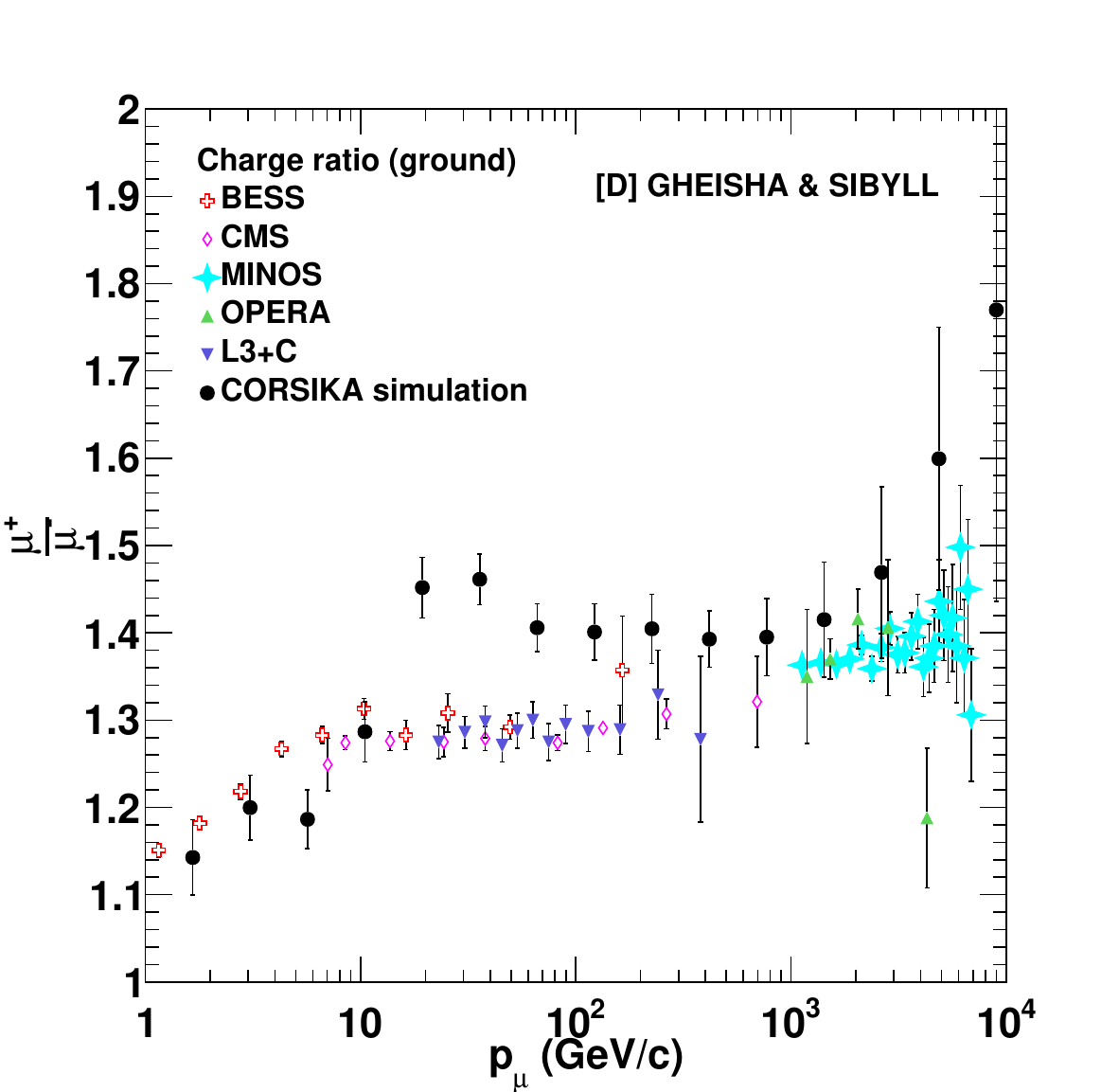}
\includegraphics[width=0.49\linewidth]{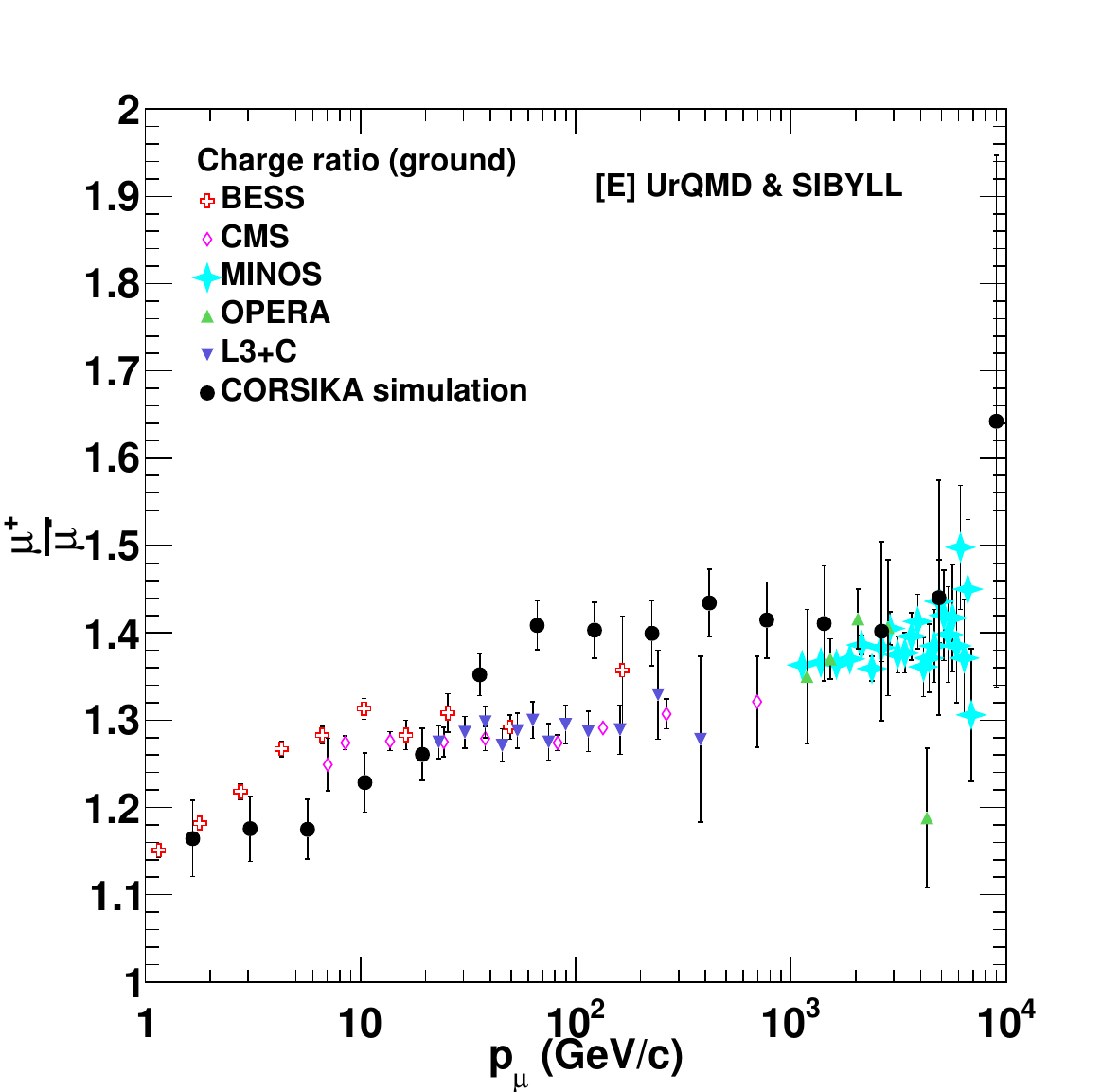}
\includegraphics[width=0.49\linewidth]{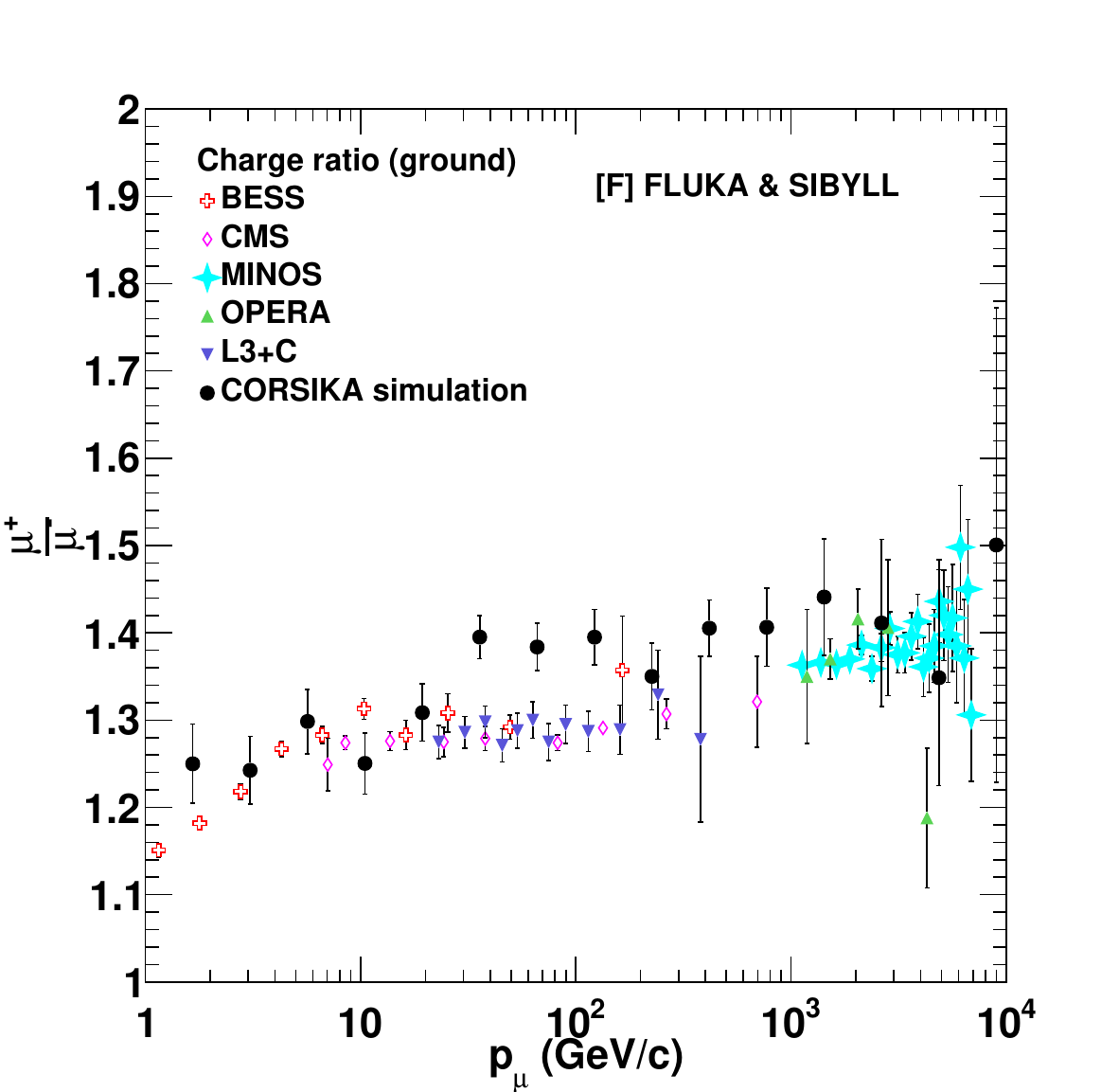}
\caption{The charge ratio $\mu^+/\mu^-$ as a function of momentum of atmospheric 
muons at the ground level calculated by CORSIKA using different
hadronic interaction model combinations along with the experimental
data collection~\cite{ParticleDataGroup:2022pth}.}
\label{figure5}	
\end{center}
\end{figure*}

Figure~\ref{figure5} shows the charge ratio $\mu^+/\mu^-$ as a function of the
momentum of atmospheric muons at the ground level calculated 
by CORSIKA using the different hadronic interaction model combinations along 
with the experimental data collection~\cite{ParticleDataGroup:2022pth}. The original 
references of experimental data are BESS~\cite{haino2004measurements}, 
CMS~\cite{CMS:2010yju}, MINOS~\cite{MINOS:2007laz}, 
OPERA~\cite{OPERA:2010cos} and L3+C~\cite{L3:2004sed}.
The jump in the momentum around 20 GeV/$c$ is due to the transition between 
low and high energy models. Muon charge ratio is more than 1 because there are more $\pi^+$ 
than $\pi^-$ produced in the shower due to the positive charge of the proton. In the higher 
momentum side, the increase in the charge ratio is due to the associated production of 
strange particles, neutral lambda and positively charged kaons which decay to positive pions 
and muons. Muon charge ratio as a function of momentum is very important to know the underlying 
physics of hadronic interactions in the atmosphere and production of various hadrons in different 
fraction and energy.
The combination of models [F] FLUKA $\&$ SIBYLL gives the best results for the muon 
charge ratio in the given momentum range as compared to other model combinations.

\begin{figure*}[h]
\begin{center}
\includegraphics[width=0.62\linewidth]{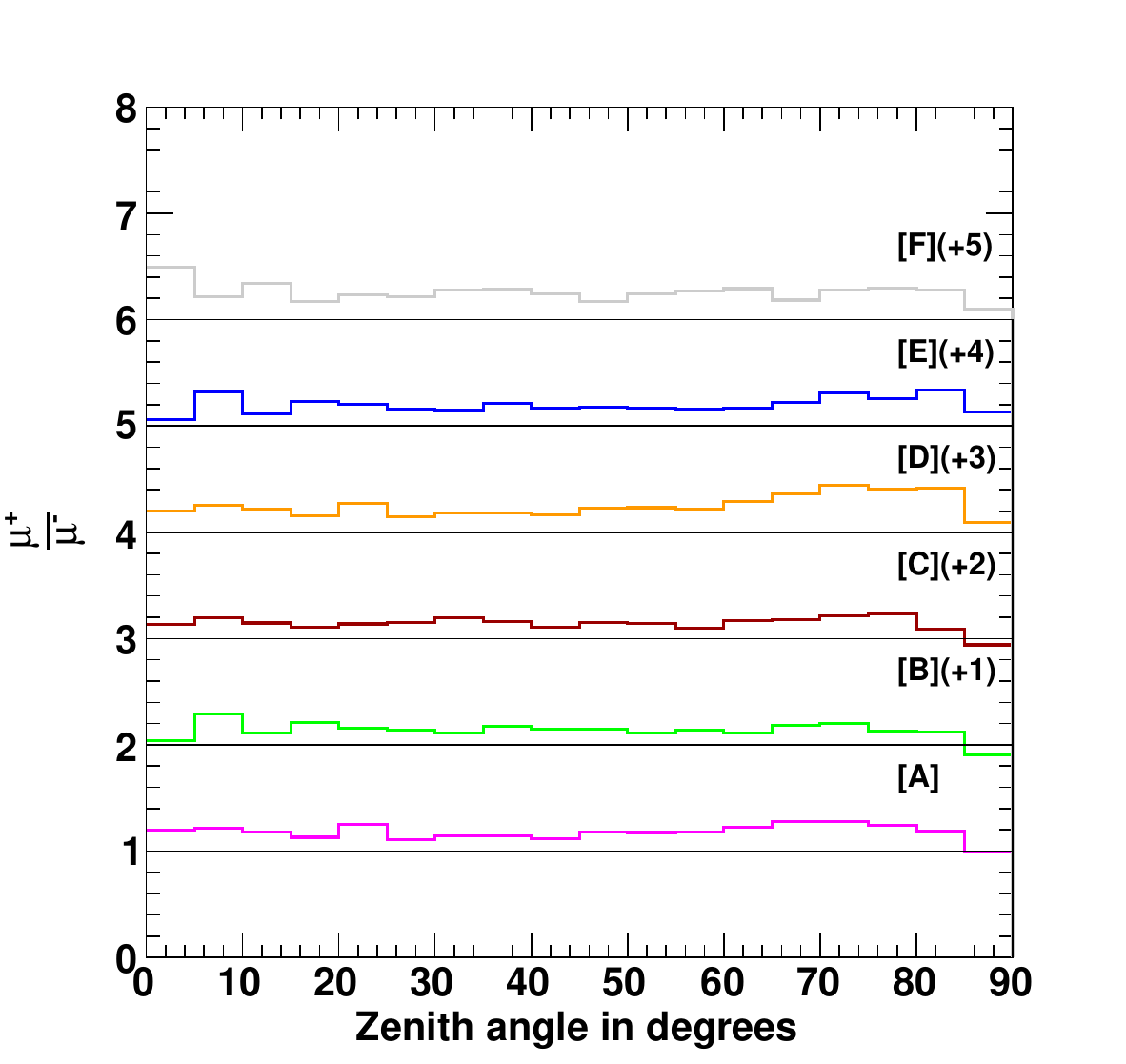}
\caption{The charge ratio $\mu^+/\mu^-$ as a function of zenith angle of 
atmospheric muons for momentum $p>0.5$ GeV/$c$ at the ground level calculated by
CORSIKA using different hadronic interaction model combinations,
[A] GHEISHA \& QGSJET, [B] UrQMD \& QGSJET, [C] UrQMD \& QGSJET-II,
[D] GHEISHA \& SIBYLL, [E] UrQMD \& SIBYLL, [F] FLUKA \& SIBYLL.
Integer numbers are added to models other than [A] as shown to separate the results. }
\label{figure6}	
\end{center}
\end{figure*}

Figure~\ref{figure6} shows the charge ratio $\mu^+/\mu^-$ as a function of zenith
angle of atmospheric muons for momentum $p>0.5$ GeV/$c$ at the ground level calculated by
CORSIKA using different hadronic interaction model combinations,
[A] GHEISHA \& QGSJET, [B] UrQMD \& QGSJET, [C] UrQMD \& QGSJET-II,
[D] GHEISHA \& SIBYLL, [E] UrQMD \& SIBYLL, [F] FLUKA \& SIBYLL.
Integer numbers are added to models other than [A] as shown to separate the results. 
Most of the models give charge ratio around 1.2 and which remains almost flat
as a function of zenith angle but slightly increases towards higher angles (60-80 degrees)
for the models [A], [C], [D] and [E]. The results can be compared with measured data.
  As the shower traverses larger path at higher zenith angles the number of hadronic interactions 
and production of particles will be different which can contribute differently in 
the muon charge ratio. The ratio increases towards higher zenith angle as the contribution
from higher energy particles which could cross the larger length. 
 It will be interesting to have muon charge measurements as a function of zenith angle.

\begin{figure*}[h]
\begin{center}
\includegraphics[width=1.1\linewidth]{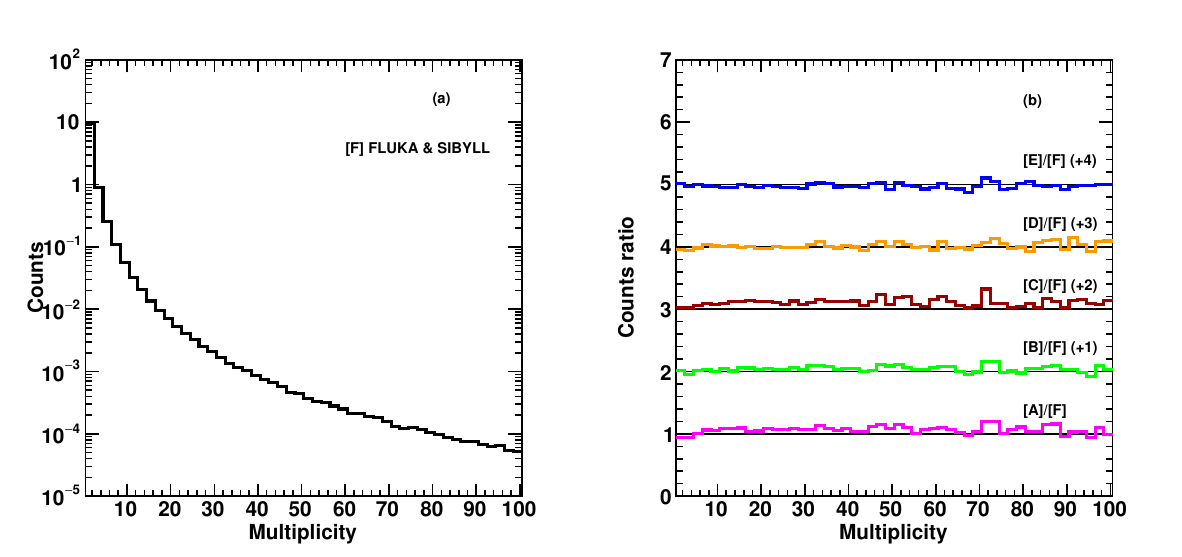}
\caption{The multiplicity of muon in a shower calculated for muon momentum $p>0.5$ GeV/$c$ using the 
(a) [F] FLUKA $\&$ SIBYLL models of CORSIKA. (b) Multiplicity ratio of model 
combinations ([A] GHEISHA \& QGSJET, [B] UrQMD \& QGSJET, [C] UrQMD \& QGSJET-II,
[D] GHEISHA \& SIBYLL and [E] UrQMD \& SIBYLL)
with respect to [F] FLUKA $\&$ SIBYLL model. An integer number is added to each model
result as shown in figure.
}
\label{figure7}	
\end{center}
\end{figure*}

Figure~\ref{figure7} shows the multiplicity of muons in a shower calculated
for muon momentum $p>0.5$ GeV/$c$ using (a) [F] FLUKA $\&$ SIBYLL model of CORSIKA. (b) Shows the
multiplicity ratios of model combinations
([A] GHEISHA \& QGSJET, [B] UrQMD \& QGSJET, [C] UrQMD \& QGSJET-II,
[D] GHEISHA \& SIBYLL and [E] UrQMD \& SIBYLL)
with respect to [F] FLUKA $\&$ SIBYLL model.
An integer number is added to each model result as shown in figure.
It can be seen from the ratios that all models give similar shape of
multiplicity distributions.
High energy model in case of [D], [E] and [F] model combination is SIBYLL.
Therefore, the multiplicity distribution is nearly same in all 3 models and
other model combinations [A], [B] and [C] show some disagreement.
At high energies or for heavier nuclei, the multiplicity might show more disagreements.
Multiplicity distributions shapes are important to get the 
extent of air shower at the ground level which will be useful to understand 
the multiple muon events in the large size detectors from each shower.

\subsection{Correlations of muons and primary particles}
 
\begin{figure*}[h]
\begin{center}
\includegraphics[width=1.1\linewidth]{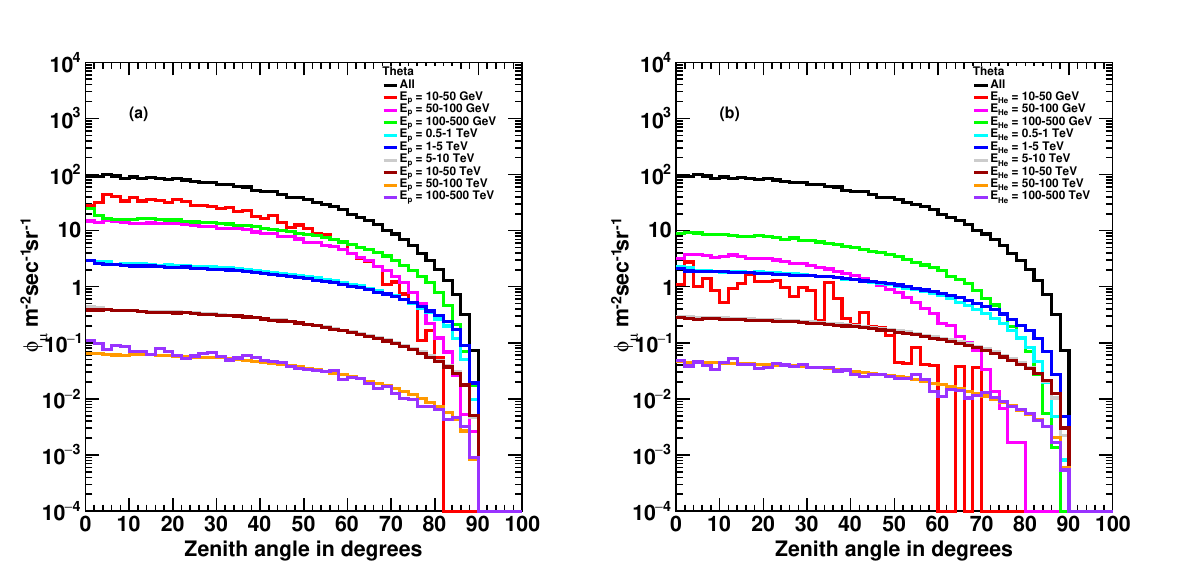}
\caption{The zenith angle distribution of muons with momentum $p>0.5$ GeV/$c$
at the ground level. Contributions in total flux are shown from various energy 
ranges for the primary particles, (a) Protons and (b) Heliums. Note that the 
energy intervals are not the same but increase with energy. The relative contribution 
of each bin is normalized as described in section 3.
}
\label{figure8}	
\end{center}
\end{figure*}

\begin{figure*}[h]
\begin{center}
\includegraphics[width=1.1\linewidth]{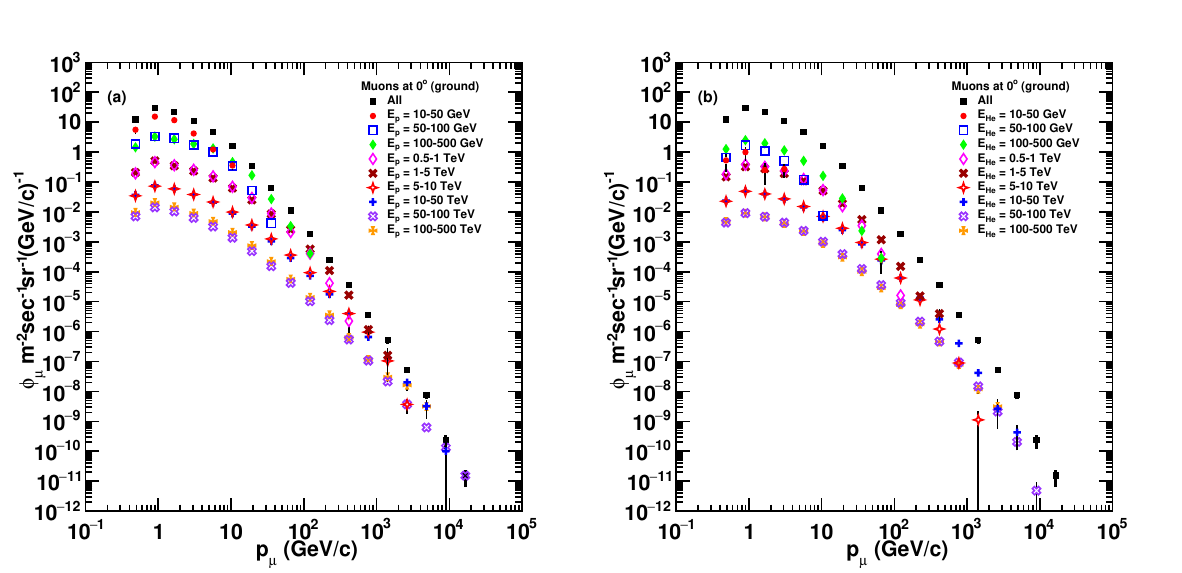}
\caption{Vertical muon flux as a function of 
 momentum  for muon momentum $p>0.5$ GeV/$c$ at the ground level.
  Contributions in the total
  flux from various energy ranges are shown for the primary 
particles, (a) Proton and (b) Helium.}
\label{figure9}	
\end{center}
\end{figure*}

\begin{figure*}[h]
\begin{center}
\includegraphics[width=1.1\linewidth]{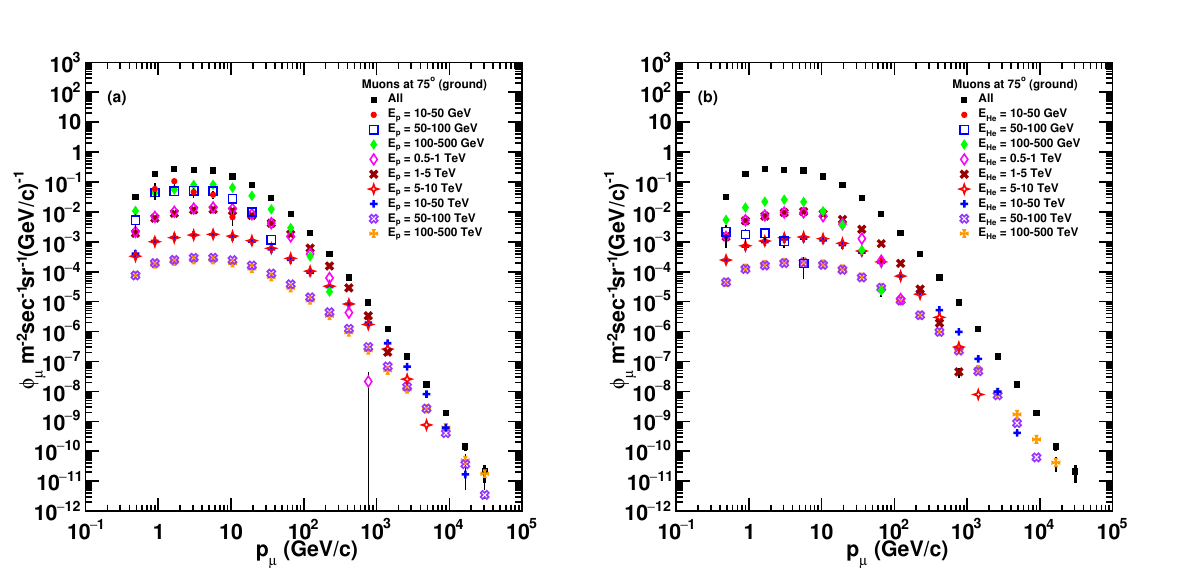}
\caption{Muon flux distribution as a function of momentum at the ground level 
for $p>0.5$ GeV/$c$ at 75$^{\circ}$ zenith angle. Contributions in the total flux 
from various energy ranges are shown for the primary particles, (a) Proton and (b) Helium.}
\label{figure10}	
\end{center}
\end{figure*}

\begin{figure*}[h]
\begin{center}
\includegraphics[width=0.49\linewidth]{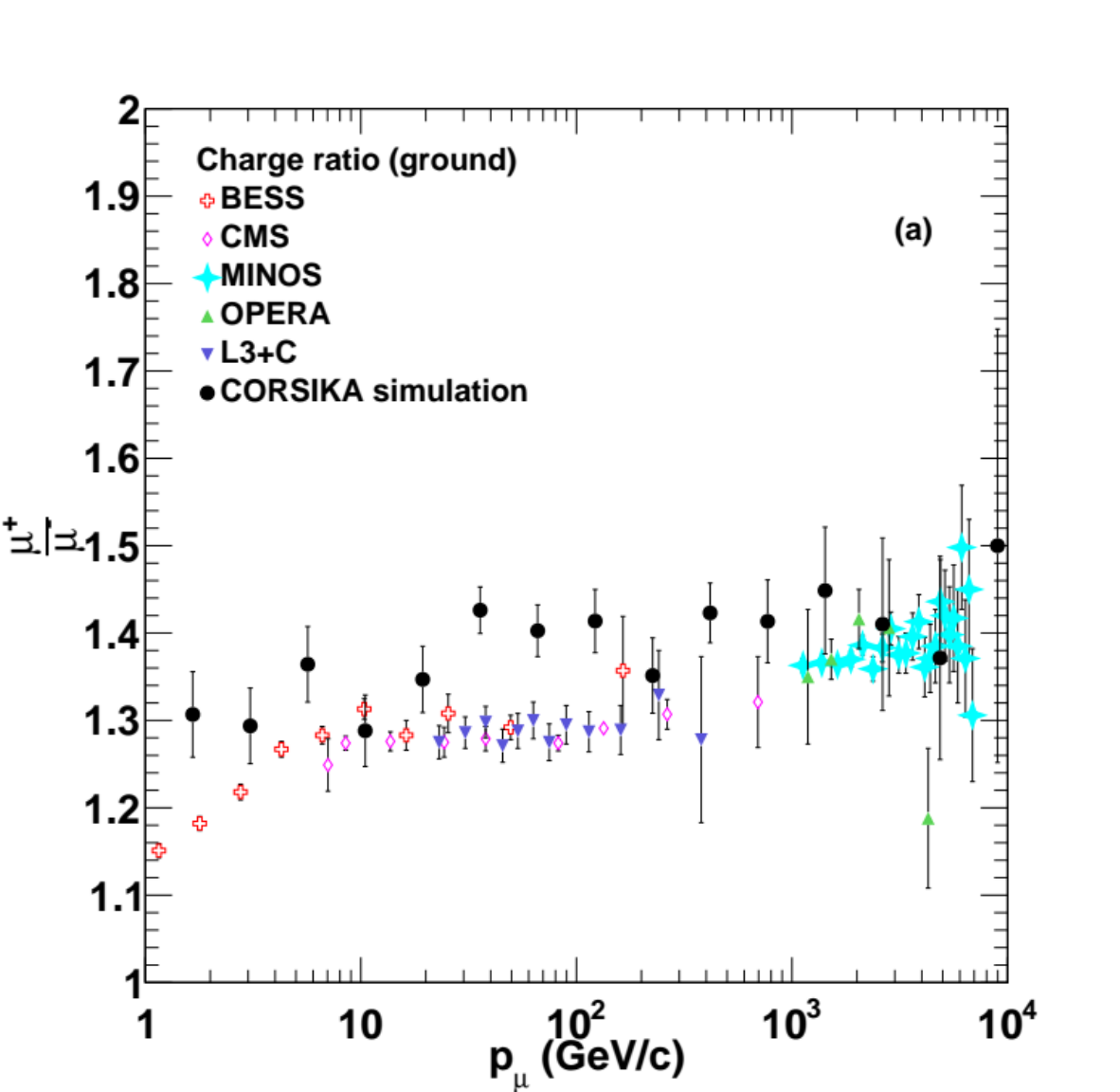}
\includegraphics[width=0.49\linewidth]{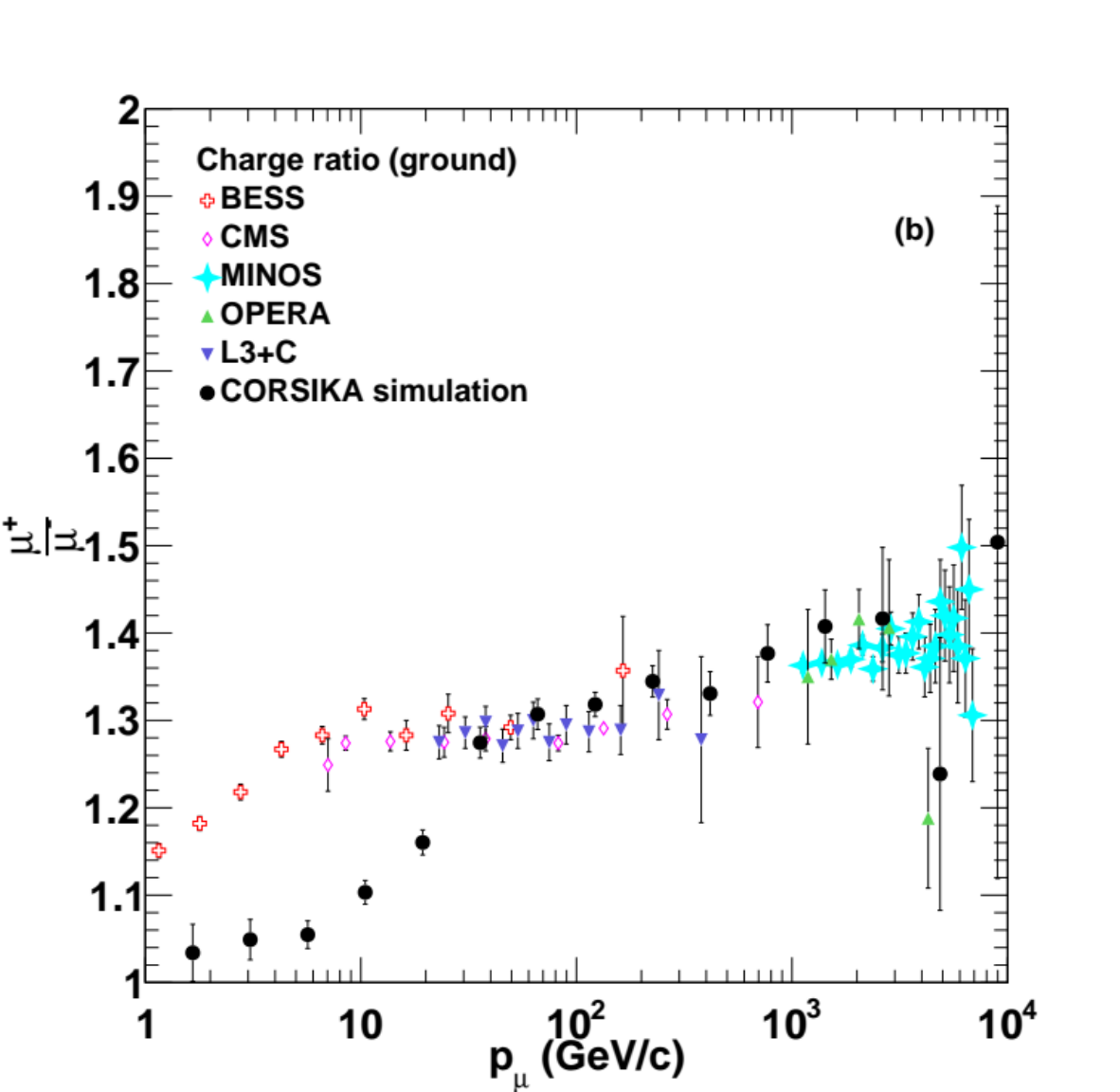}
\caption{The charge ratio $\mu^+/\mu^-$ as a function of momentum of atmospheric 
muons at the ground level calculated by CORSIKA for (a) Proton and (b) Helium using 
hadronic interaction model combinations [F] along with the experimental
data collection~\cite{ParticleDataGroup:2022pth}.}
\label{figure11}	
\end{center}
\end{figure*}

\begin{figure*}[h]
\begin{center}
\includegraphics[width=1.1\linewidth]{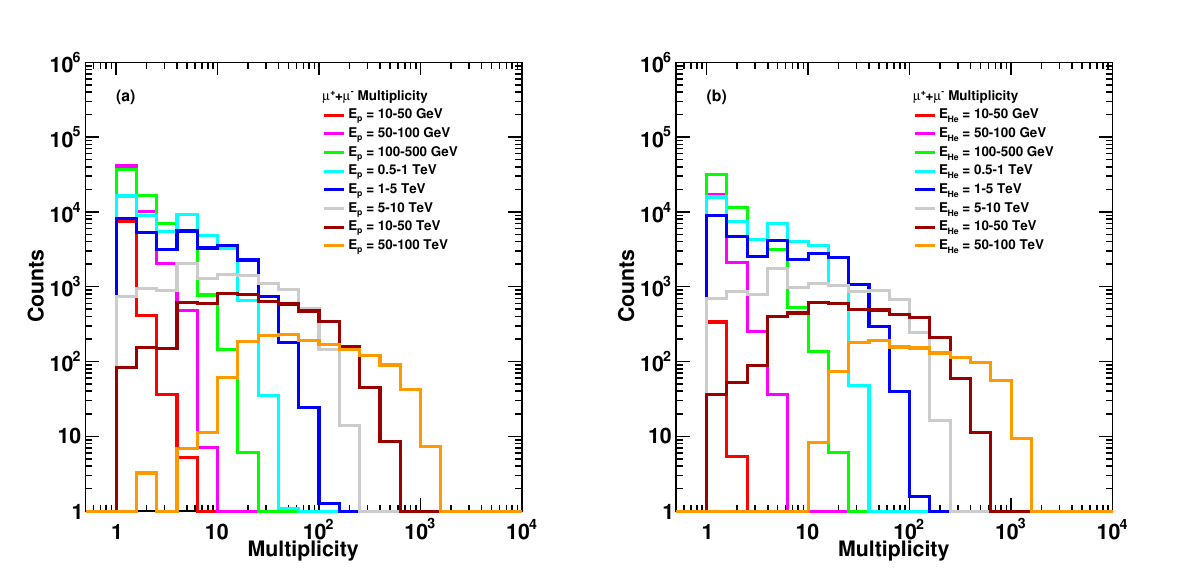}
\caption{Multiplicity variation of cosmic muons above momentum $p>0.5$ GeV/$c$
 from various energy ranges of the primary particles,
 (a) Protons and (b) Heliums. Each energy interval corresponds to 0.1 million
 events of primaries.}
\label{figure12}	
\end{center}
\end{figure*}

 Contributions in the different muon observables coming from different
energy as well as zenith angle intervals of primary protons and heliums
are obtained. 
The primary particles and their energy dependence in the muon 
distributions at ground level gives a deep understanding of the air-showers 
generation in the atmosphere.
Figures~\ref{figure8} to~\ref{figure12} are calculated by CORSIKA using 
[F] SIBYLL $\&$ FLUKA model. 

Figure~\ref{figure8} shows the zenith angle distribution of muons with momentum
$p>0.5$ GeV/$c$ at the ground level. Contributions in total flux are shown
from various energy ranges
for the primary particles, (a) Protons and (b) Heliums. Note that
the energy intervals are not the same but increase with energy.
The energy bins are normalized as described in section 3.
The lower zenith angles get more contribution from the lower energies 
of primary particles while the higher angles get more contribution from the higher 
energies of primary particles. The bulk contribution from Helium contributes
to higher energy bin as compared to that for proton.

Figure~\ref{figure9} shows the vertical muon flux 
as a function of momentum at the ground level for muon momentum $p>0.5$ GeV/$c$. 
Contributions in the total flux from various energy ranges are shown for the primary particles,
(a) Protons and (b) Heliums.
Note that the energy intervals are not the same but increase with energy but they
are normalized as per the power law distribution. 
Contribution to lower energy of muons distribution comes from lower energy
part of primary particles.  
The higher energy part of primary particles give contribution to the higher
energy part of the muon distribution and thus there is a crossover between
distributions of two successive energy histograms of primaries.

Figure~\ref{figure10} shows the muon flux distribution as a function of momentum 
for $p>0.5$ GeV/$c$ at 75$^{\circ}$ zenith angle. Contributions in the total flux 
from various energy ranges are shown for the primary particles, (a) Protons and (b) Heliums.
One can notice that the contribution from higher energy bins [100-500] GeV is more 
dominant as compared to the lower energy bins. It shows that the muons at higher 
zenith angles get contributions from higher energy primary particles. 
One can compare it with Figure~\ref{figure9}, where the lower energy bins
were more dominating in the lower energy part of muon distribution.
%The momentum distribution at higher zenith angles are flater in the 
%momentum range from 1 to 10 GeV/$c$ and this flat region is more for the helium as 
%compared to the proton.

Figure~\ref{figure11} shows the charge ratio $\mu^+/\mu^-$ as a function of the
momentum of atmospheric muons at the ground level calculated 
by CORSIKA separately for (a) Proton and (b) Helium using hadronic interaction model [F] combinations along 
with the experimental data collection~\cite{ParticleDataGroup:2022pth}. The original 
references of experimental data are BESS~\cite{haino2004measurements}, 
CMS~\cite{CMS:2010yju}, MINOS~\cite{MINOS:2007laz}, 
OPERA~\cite{OPERA:2010cos} and L3+C~\cite{L3:2004sed}.
The weight of Helium in the spectra is less than 10$\%$.
The figure shows that charge ratio would decrease with the increase in Z of the primary 
particles. But due to the decreasing flux of higher Z particles their weight 
to the spectra of charge ratio will be very small.

Figure~\ref{figure12} shows the multiplicity variation of muons 
above momentum $p>0.5$ GeV/$c$ from various energy ranges of the primary particles,
(a) Protons and (b) Heliums.
Each energy interval corresponds to 0.1 million events of primaries. 
The lower multiplicity muon events are coming mainly from the lower energy 
primaries and higher multiplicity events are coming from the higher energy primaries. 
Helium contributes more towards higher multiplicity.
This study is useful to correlate the muon multiplicity at the ground with the
primary particle energy.

\begin{figure*}[h]
\begin{center}
\includegraphics[width=1.1\linewidth]{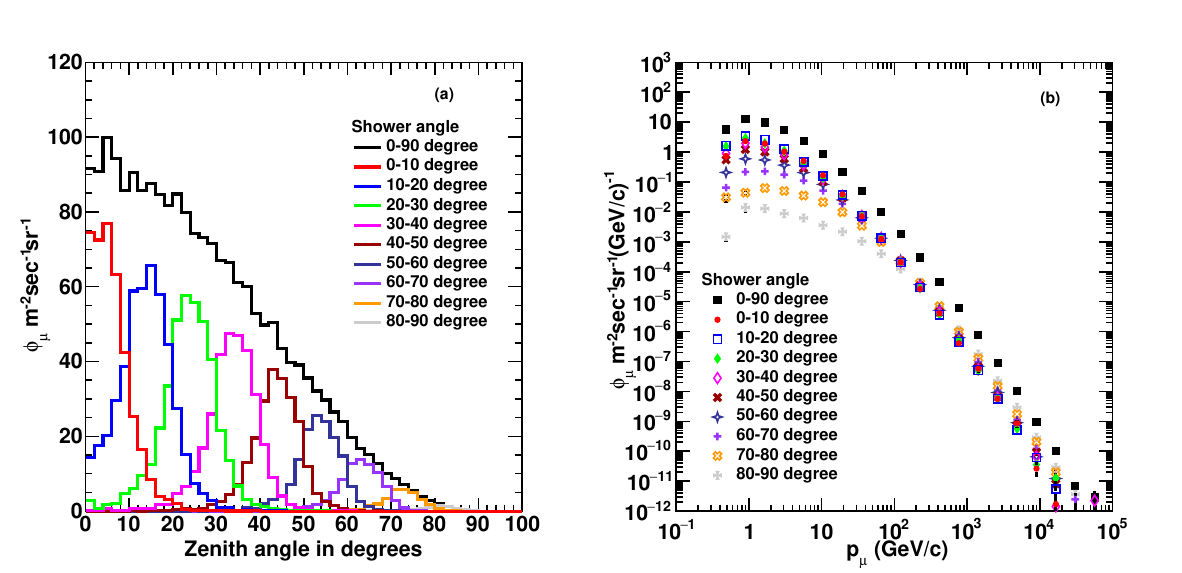}
\caption{Muon flux distributions at ground level for muon momentum $p>0.5$ GeV/$c$. 
  The contributions of primary particles from different angle intervals are shown in
  (a) Zenith angle distribution of muons
and (b) Momentum distribution of muons.}
\label{figure13}	
\end{center}
\end{figure*}

Figure~\ref{figure13} shows the muon
flux distributions at ground level for muon momentum $p>0.5$ GeV/$c$. 
  The contribution of primary particles from different angle intervals are shown in
  (a) Zenith angle distribution of muons
and (b) Momentum distribution of muons.
It is clear from Figure~\ref{figure13}(a) that the secondary
particles (muons in this case) keep the memory of the direction
of the primary particles and Figure~\ref{figure13}(b) shows that
the contribution to the lower energy part of the muon distributions
comes from primary particles which are at lower zenith angles. 
The lower momentum muon flux gets more contributions from primary zenith angle between 
10 to 20 degrees as compared to 0 to 10 degrees which shows the increase in the muon production 
with the higher path-lengths in the atmosphere. As the shower zenith angles increases
further the muon survival probablity also decreases. The higher momentum muons get
contribution mainly 
from higher shower zenith angles which shows the penetration of high energy primary 
particles and secondary pions, kaons and other hadrons deep into the atmosphere and 
producing high energy muons near the Earth's surface. Contribution in the lower momentum 
muons from 0.5 GeV/$c$ to 5 GeV/$c$ is mainly from shower angle from 0 to 30 degrees. 
Contribution in the muon momentum region between $10^2$ GeV/$c$ to $10^3$ GeV/$c$
is nearly equal from all the shower angles and the contribution in $10^3$ GeV/$c$ to $10^4$ GeV/$c$
region is dominantly from showers between 70 to 90 degree.

The explanations to Figure~\ref{figure8} to Figure~\ref{figure13} are easy to
understand and give us a quantitative understanding of correlation between
the primary particle direction and energy with muon direction and energy.

\subsection{Flux and ratios of various particles}

\begin{figure*}[h]
\begin{center}
\includegraphics[width=1.1\linewidth]{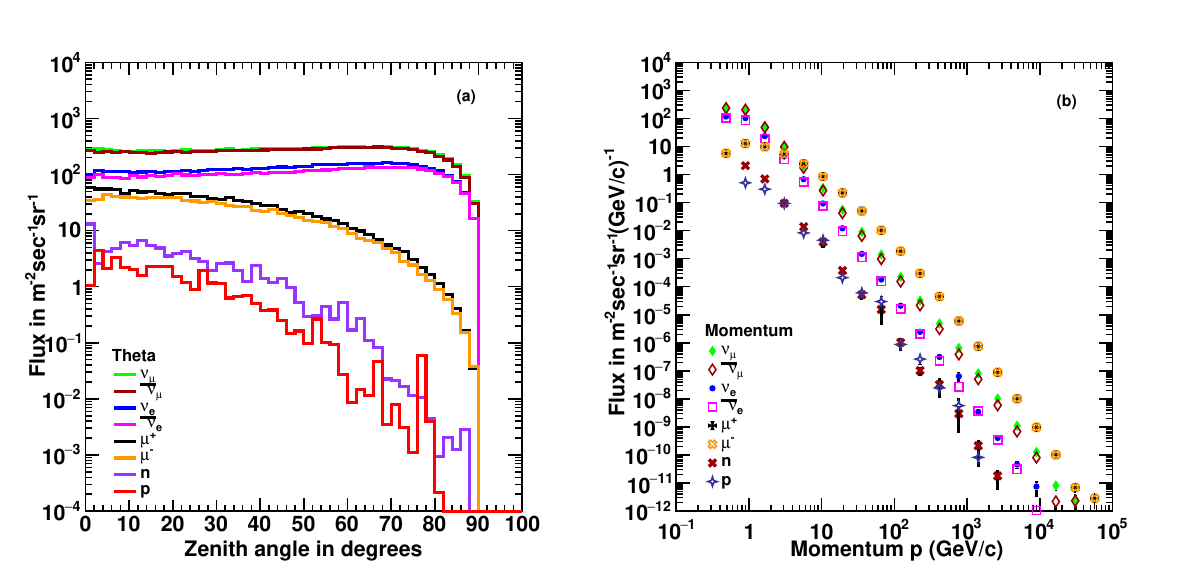}
\caption{The flux of dominant cosmic ray shower particles reaching 
at ground level for particle momentum $p>0.5$ GeV/$c$ as a function of (a) Zenith angle and 
(b) Momentum of $\nu_{\mu}$, $\bar{\nu_{\mu}}$, $\nu_{e}$, 
$\bar{\nu_{e}}$, $\mu^{+}$, $\mu^{-}$, $n$ and $p$.}
\label{figure14}	
\end{center}
\end{figure*}

\begin{figure*}[h]
\begin{center}
\includegraphics[width=1.1\linewidth]{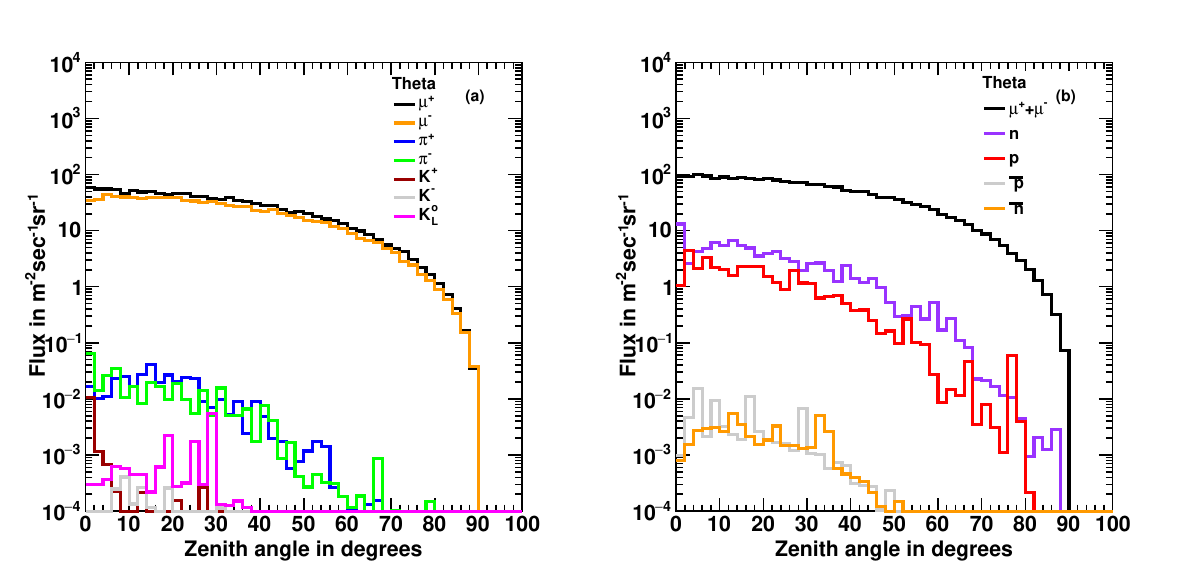}
\caption{Flux of the various particles as a function of the zenith angle from 
  the cosmic ray shower reaching at ground level for particle momentum $p>0.5$ GeV/$c$:
  (a) $\mu^{+}$, $\mu^{-}$, $\pi^{+}$, $\pi^{-}$, $K^{+}$,
$K^{-}$ and $K^{\circ}_{L}$ and
(b) $\mu^{+}$+$\mu^{-}$, $n$, $p$, $\bar{p}$ and $\bar{n}$.}
\label{figure15}	
\end{center}
\end{figure*}

\begin{figure*}[h]
\begin{center}
\includegraphics[width=1.1\linewidth]{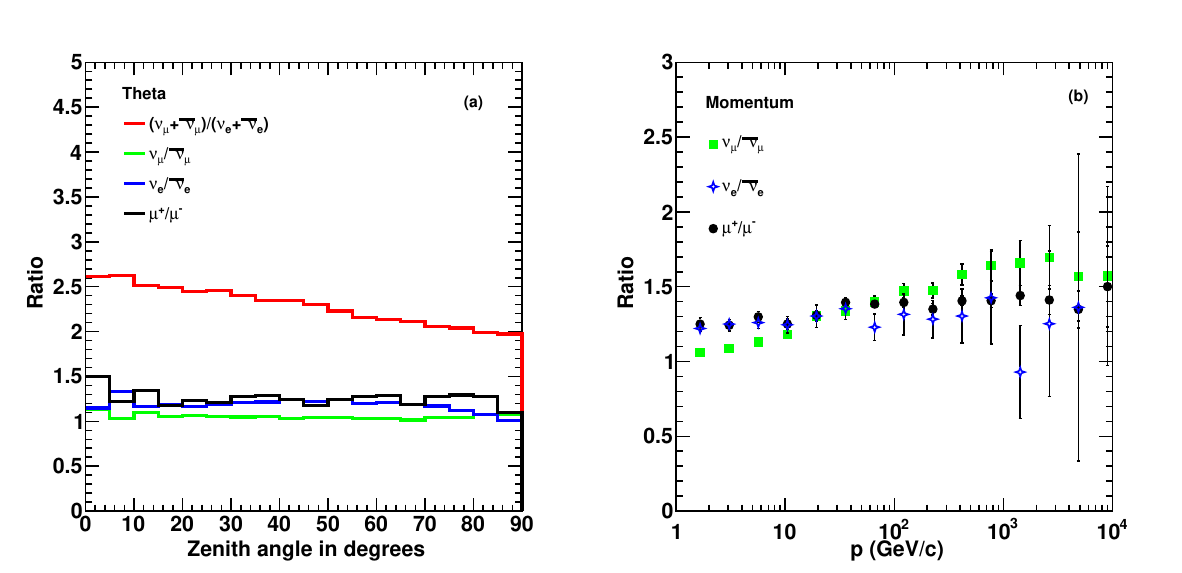}
\caption{Ratio of various leptons for momentum $p>0.5$ GeV/$c$ at ground level
  (a) Zenith angle distribution and (b) Momentum distribution.}
\label{figure16}	
\end{center}
\end{figure*}

\begin{figure*}[h]
\begin{center}
\includegraphics[width=0.6\linewidth]{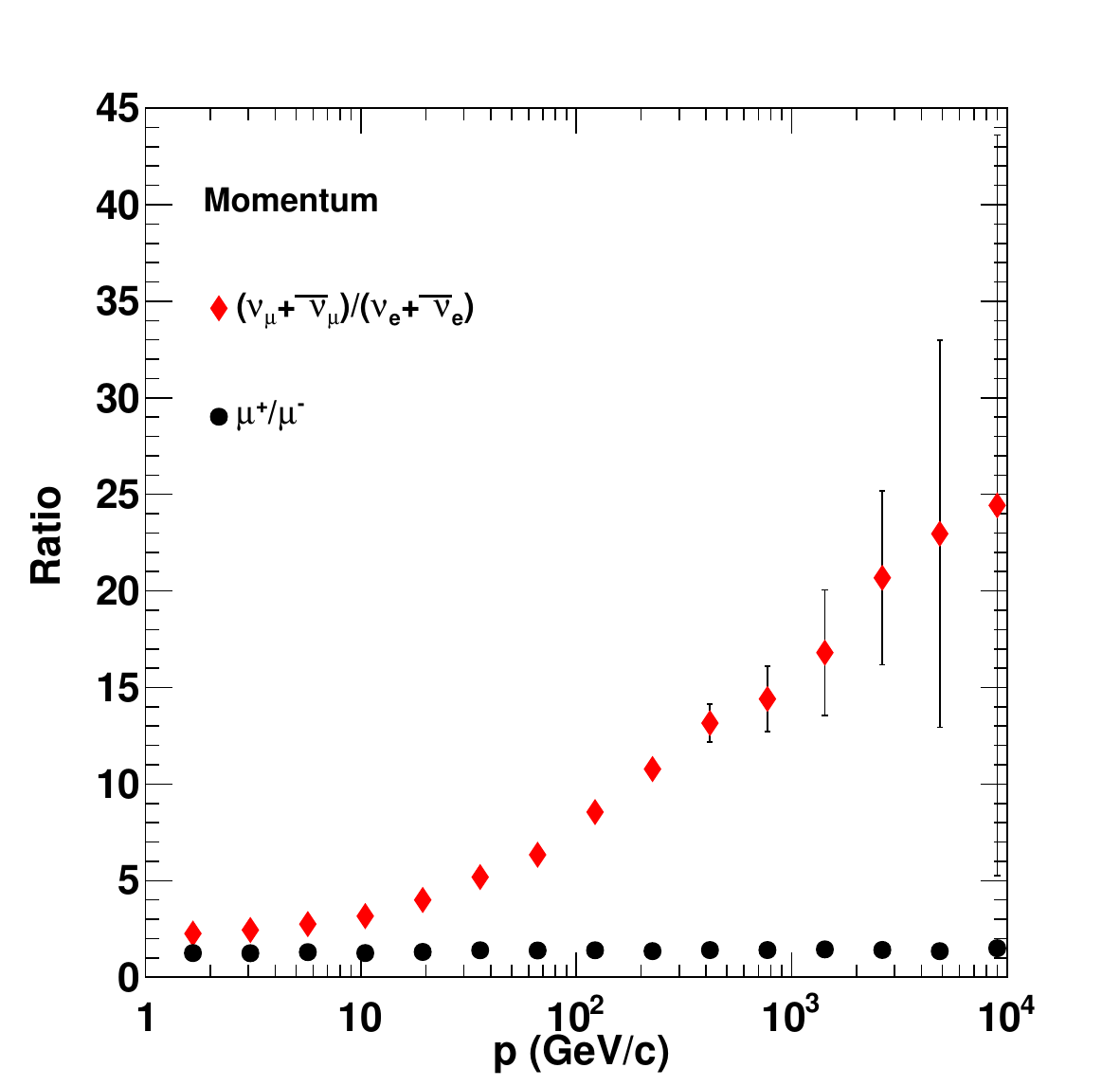}
\caption{Ratio of numbers of various leptons for momentum $p>0.5$ GeV/$c$ at ground level
  (a) Zenith angle distribution and (b) Momentum distribution.}
\label{figure17}	
\end{center}
\end{figure*}

\begin{figure*}[h]
\begin{center}
\includegraphics[width=1.1\linewidth]{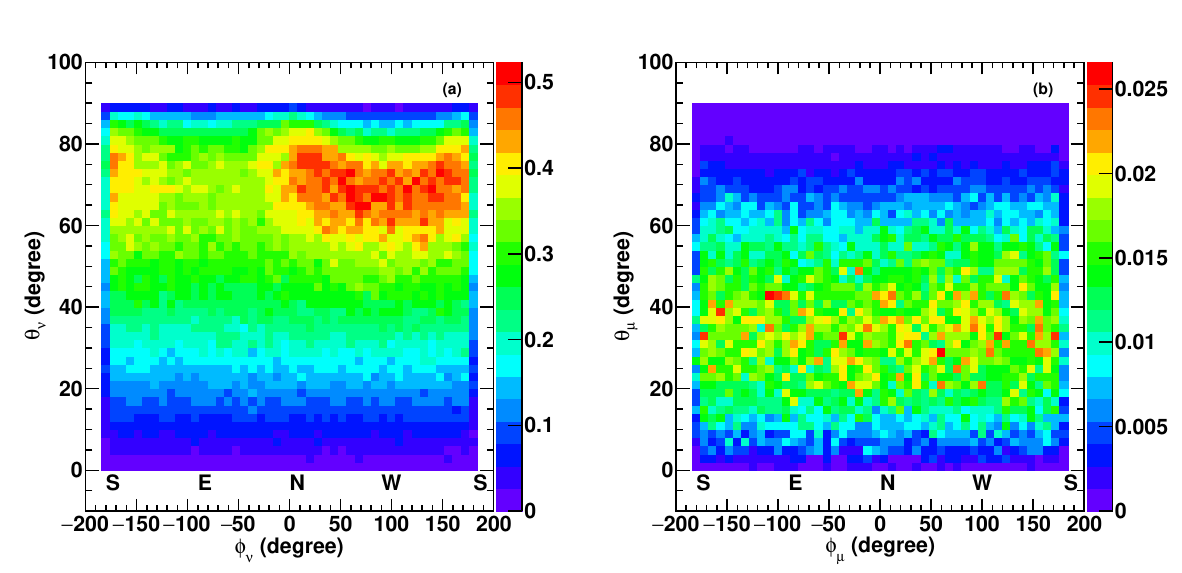}
\caption{East-west asymmetry at ground level for particle momentum $p$>0.5 GeV/$c$
  for (a) Neutrinos ($\nu_\mu$, $\bar{\nu_\mu}$, $\nu_e$ and $\bar{\nu_e}$) 
and (b) Muons ($\mu^{+}$ and $\mu^{-}$).}
\label{figure18}	
\end{center}
\end{figure*}

We also looked at the distributions of other particles and their contributions
at ground except the electromagnetic part.
Figures~\ref{figure14} to~\ref{figure18} are calculated by CORSIKA using 
[F] SIBYLL $\&$ FLUKA model. 
Figure~\ref{figure14} shows the flux of the dominant cosmic ray
shower particles reaching at ground level for particle momentum $p>0.5$ GeV/$c$
as a function of (a) zenith angle and (b) momentum of $\nu_\mu$, $\bar{\nu}_\mu$, $\nu_e$, 
$\bar{\nu}_e$, $\mu^{+}$, $\mu^{-}$, $n$ and $p$.
The figure in (a) shows that although the muon flux falls with increasing angle,
the neutrino flux remains flat and then increases towards higher angles.
Neutrons and protons remain the third dominant component at the ground after
neutrinos and muons.
The results of Fig.~\ref{figure14}(a) can be analyzed as follows:
The muon neutrinos are produced in the production as well as in the decay of muons
while the electron neutrinos are only produced in the decay of muons  
giving nearly twice muon neutrinos over the electron neutrinos.
More muons are produced and decay at the higher zenith angles giving more and more
neutrinos at the higher zenith angles. Muon flux at higher 
zenith angles is lower because of muon decay.
Proton distribution is below neutron as they lose energy during propagation in
the atmosphere. Similarly, the results of Fig.~\ref{figure14}(b) can be analyzed
as follows: The flux of muon-neutrinos in the low momentum region from
0.5 GeV/$c$ to 5 GeV/$c$ are nearly twice of the electron neutrino while towards the
higher momentum side this ratio increases up to 10 to 20 times because higher
momentum muons do not decay to electron neutrinos.
Proton flux at higher momentum is similar to the neutron 
flux but there is significant difference in the lower momentum side due to the 
proton energy loss during its propagation in the atmosphere.

Figure~\ref{figure15} shows the flux of the various particles as a function 
of the zenith angle from the cosmic ray shower reaching at ground level for 
particle momentum $p>0.5$ GeV/$c$:
(a) $\mu^{+}$, $\mu^{-}$, $\pi^{+}$, $\pi^{-}$, $K^{+}$,
$K^{-}$ and $K^{\circ}_{L}$ and
(b) $\mu^{+}$+$\mu^{-}$, $n$, $p$, $\bar{p}$ and $\bar{n}$. 
These distributions are important and if compared with the measurements will
help in validation of models. 
After muons and nucleons,
pions are the most abundant particles at the ground level.

Table~\ref{table6} shows the integrated particle flux from zenith angle 
distribution at ground level for particle momentum $p>0.5$ GeV/$c$.
\begin{table*}[h]
\begin{center}
\caption{Integrated particles flux obtained from zenith angle distribution at ground level 
 for particle momentum $p>0.5$ GeV/$c$.}
\label{table6}
\resizebox{\columnwidth}{!}{%
\begin{tabular}{c c c c} 
\hline	
\textbf{Particles (at G.L)}	 & \textbf{Area of $\theta$ distribution} & \textbf{Particles (at G.L)} & \textbf{Area of $\theta$ distribution} \\
  & (m$^{-2}$sec$^{-1}$) &  & (m$^{-2}$sec$^{-1}$)\\
\hline
$\nu_\mu$ & 1657.97 & $\pi^{+}$ & 0.0237 \\ 
$\bar{\nu}_\mu$ & 1590.28 & $\pi^{-}$ & 0.0178 \\ 
$\nu_e$ & 802.83 & $\bar{p}$ & 0.0040 \\ 
$\bar{\nu}_e$ & 679.52 & $\bar{n}$ & 0.0034\\
$\mu^{+}$ & 106.12 & $K^{+}$ & 0.0001 \\ 
$\mu^{-}$ & 85.07 & $K^{-}$ & 0.0001 \\ 
$n$ & 6.07 & $K^{\circ}_{L}$ & 0.0012 \\ 
$p$ & 2.26 & $K^{\circ}_{S}$ & 8.2244$\times$10$^{\rm -7}$ \\ 
\hline
\end{tabular}
}
\end{center}
\end{table*}

Figure~\ref{figure16} show the ratio of numbers 
of various leptons at ground level for momentum $p$>0.5 GeV/$c$ as
(a) zenith angle distribution and (b) momentum distribution.
The $\nu_\mu/\bar{\nu}_\mu$ ratio remains flat over angle. The ratio
$\nu_e/\bar{\nu}_e$ is similar to the ratio $\mu^+/\mu^-$.
The ratio $(\nu_\mu+\bar{\nu}_\mu)/(\nu_e+\bar{\nu}_e)$ decreases with
increasing zenith angle since muons at higher zenith angles have more
probability to decay giving more number of electron neutrinos.

The measurement of ratio $\mu^+/\mu^-$ can be connected to the neutrino ratios
in both the flavours ($e$ and $\mu$) as we explain in the following. 
The decay of positive and negative particles (which depends on the energy and
height) are given by \\
\ \\
$\pi^+ \rightarrow \mu^+ + \nu_\mu$ and $\mu^+ \rightarrow e^+ + \bar{\nu}_\mu + \nu_e$, \\
$\pi^- \rightarrow \mu^- + \bar{\nu}_\mu$ and $\mu^- \rightarrow e^- + {\nu}_\mu + \bar{\nu}_e$. \\
\ \\
The charge ratio of survived muons at ground level will be equal to 
the charge ratio of decayed muons and hence would also be equal to $\nu_e$/$\bar{\nu}_e$.     
Deviations of $\nu_e$/$\bar{\nu}_e$ from the muon charge ratio at ground level would require 
further understanding of phenomenological models and experimental data for the bulk of 
the air showers at ground level.
Muon charge ratio is equal in the survived and total samples.
At ground level, deviations of $\nu_\mu$/$\bar{\nu}_\mu$ from the muon charge ratio mainly
comes from the opposite contributions in muon neutrinos from muon decays. 
Now assume if there are more $\mu^+$ than $\mu^-$ detected at ground level and if 
muon survival probability was more (the case of lower zenith angle and higher 
momentum muons) then there will be more associated production of $\nu_\mu$ than 
$\bar{\nu}_\mu$.
In the case of higher zenith angle and lower momentum muons, the 
decay probability increases and hence there will be more $\nu_e$ than $\bar{\nu}_e$ 
and $\bar{\nu}_\mu$ than $\nu_\mu$.

 Figure~\ref{figure17} shows the ratio of numbers 
of various leptons at ground level for momentum $p$>0.5 GeV/$c$ as
a function of zenith angle.
The numerator $\nu_\mu+\bar{\nu}_\mu$ are produced in both the muon production 
and decay while $\nu_e+\bar{\nu}_e$ are produced only in the muon decays. 
The ratio $(\nu_\mu+\bar{\nu}_\mu)/(\nu_e+\bar{\nu}_e)$ gives the estimation of muon decays 
and hence decreases with increasing zenith angle and increases with increasing energy.

Figure~\ref{figure18} shows the East-west asymmetry at ground level for 
particle momentum $p$>0.5 GeV/$c$ for (a) Neutrinos ($\nu_\mu$, $\bar{\nu_\mu}$, $\nu_e$ 
and $\bar{\nu_e}$) and (b) Muons ($\mu^{+}$ and $\mu^{-}$). Asymmetry is 
visible in figure (a) of neutrinos at higher zenith angles while it is not clearly 
seen in the case of figure (b) of muons. 
At the ground, more particles will reach from the west as compared to east direction 
because the shower of the secondary particles follows the direction of the primary 
cosmic particles (proton and alpha etc). These primaries are positively charged and 
hence are deflected by Earth's magnetic field in the west direction.
It can be readily seen that east-west asymmetry is visible at higher 
zenith angles of the neutrinos and not visible for the muons as they lose energies in 
the atmosphere during propagation. Asymmetry increase at higher altitude and decrease 
with increasing latitude. It also depends on momentum cut on the particles.
%It was measured by Johnson and Street in 1933. 
%This effect will be more pronounced near the equator.

\section{Conclusions}

We presented a comprehensive study of air shower simulations using various
combinations of hadronic models in the CORSIKA package. Six model combinations are
used one from the low energy models GHEISHA, UrQMD and FLUKA and one from the
high-energy models QGSJET, QGSJET-II and SIBYLL.
In this study, we focus on the bulk of the particles at ground level
and obtain the absolute yields using normalized simulation results.
We have used data for muon distributions namely, zenith angle distribution,
momentum distributions (at $0^\circ$ and $75^\circ$) and muon charge ratios
and recent parametrizations to compare with six model combinations.
These distributions from various models are quantified using the $\chi^{2}$ method
and integrated flux obtained from all the distributions are compared
among the models. 
All the model combinations produce the experimental shape of zenith angle
distribution and underestimate the flux at very high zenith angles.   
Overall, GHEISHA and FLUKA give the best results for zenith angle distribution. 
All model combinations give a good description of momentum distribution although
in the low energy region, they underestimate the flux.
Overall the model combinations, UrQMD \& SIBYLL  and FLUKA \& SIBYLL give best
results.
The charge ratio of muons as a function of the momentum is obtained 
with the different hadronic models and compared with the experimental data.
The combination of models FLUKA \& SIBYLL gives the best results for the
muon charge ratio in the given momentum range as compared to other model combinations.
The zenith angle dependence of the charge ratio of muons is predicted and compared
among different models. The multiplicity distributions from different models
have been obtained. The differences in the results among various models have
been outlined. 

 Contribution in the different muon observables coming from different energies as
well as zenith angle intervals of primary protons and heliums is obtained.
Higher zenith angle gets more contribution from higher energies of primary
particles. The muons at ground keep the memory of the direction of the primary
particles.

We predict the zenith angle and momentum distribution of all other
particles $\nu_{\mu}$, $\bar{\nu_{\mu}}$, $\nu_{e}$, $\bar{\nu_{e}}$, $\mu^{+}$,
$\mu^{-}$, $n$, $p$, $\pi$ and $K$ and their ratios which could be
measurable at ground in addition to muons.
Neutrons and protons remain the third dominant component at the
ground after neutrinos and muons.
 These distributions are important and if compared with the measurements will
lead to the validation of models.
Various leptonic ratios are obtained as a function of zenith angle and momentum.
These are important for a quantitative understanding of fluxes of muons and
neutrino flavours and their correlations which is an important prerequisite
for neutrino oscillation physics. 
Distribution of muons and neutrinos as a function of
$\theta$ and $\phi$ are presented to quantify East-west asymmetry.

\section*{Acknowledgements}
The authors acknowledge fruitful discussions and help in simulations from
Dr. Vineet Kumar, Mr. Raman Sehgal and Sundaresh Sankrith.


\begin{thebibliography}{100}
\medskip

\bibitem{grieder444507108cosmic} P. K. F. Grieder, Cosmic rays at earth: Researcher's reference manual and data book, Elsevier (2001).

\bibitem{Sogarwal:2022kgw}
H.~Sogarwal and P.~Shukla, JCAP \textbf{07} 11, (2022).
   
\bibitem{ParticleDataGroup:2022pth}
R.~L.~Workman \textit{et al.} [Particle Data Group], PTEP \textbf{083C01} (2022).

\bibitem{Dembinski:2020dam}
H.~P.~Dembinski, R.~Ulrich and T.~Pierog, "Future Proton-Oxygen Beam Collisions at the LHC for Air Shower Physics", PoS \textbf{ICRC2019}, 235 (2020).


%%%%%%%%%%%%%%%%%%%%%%%%%%%%%%%%%%%%%%%%%%%%%%
\bibitem{Heck:1998vt} D.~Heck, G.~Schatz, T.~Thouw, J.~Knapp, J.~N.~Capdevielle, CORSIKA: A Monte Carlo code to simulate extensive air showers,  \textbf{FZKA-6019} (1998).

\bibitem{userguide} D.~Heck, T.~Pierog, Extensive air shower simulation with CORSIKA: A users guide (Version 7.7100) (2016).

\bibitem{QGS1} N.~N.~Kalmykov, and S.~S.~Ostapchenko Phys. Atom. Nucl., 56, 346. Yad. Fiz. 56N3 (1993).

\bibitem{QGS2} N.~N.~Kalmykov, S.~S.~Ostapchenko and A.~I.~Pavlov, (1994), Bull. Russ. Acad. Sci. Phys., 58, (1966).
  
\bibitem{QGS3} N.~N.~Kalmykov, S.~S.~Ostapchenko and A.~I.~Pavlov, Nucl. Phys. Proc. Suppl., \textbf{52}, 17 (1997).

\bibitem{QGSII} S.~Ostapchenko, Phys. Rev. D \textbf{83}, 014018 (2011). 

\bibitem{QGSII1} S.~Ostapchenko, Phys. Rev. D \textbf{89}, 074009 (2014).

\bibitem{fletcher1994s} R.~S.~Fletcher, T.~K.~Gaisser, P.~Lipari, and T.~Stanev, Phys. Rev. D \textbf{50}, 5710 (1994).

\bibitem{sibyll} J.~Engel, T.~K.~Gaisser, T.~Stanev and P.~Lipari, 1992, Phys. Rev. D \textbf{46}, 5013 (1992).

\bibitem{ahn2009cosmic} Ahn, Eun-Joo, R.~Engel, T.~K.~Gaisser, P.~Lipari and T.~Stanev, Phys. Rev. D \textbf{80},  094003 (2009).
 
\bibitem{riehn2015new} F.~Riehn, R.~Engel, A.~Fedynitch, T.~K.~Gaisser and T.~Stanev, PoS, \textbf{ICRC2015}, 558 (2016).
  
\bibitem{riehn2015hadronic} F.~Riehn, Hadronic multiparticle production with Sibyll, KIT, Karlsruhe, Dept. Phys. https://publikationen.bibliothek.kit.edu/1000052699 (2015).
                    
\bibitem{Gheisha} H.~Fesefeldt, The simulation of hadronic showers: Physics and Applications, PITHA-85-02 (1985).

\bibitem{bass1998microscopic} S.~A.~Bass, {\it et. al}, Prog. Part. Nucl. Phys., \textbf{41}, 255 (1998). 
        
\bibitem{bleicher1999relativistic} M.~Bleicher, {\it et. al}, J. Phys. G \textbf{25}, 1859 (1999). 

\bibitem{Bohlen:2014buj}
T.~T.~B\"ohlen, F.~Cerutti, M.~P.~W.~Chin, A.~Fass\`o, A.~Ferrari, P.~G.~Ortega, A.~Mairani, P.~R.~Sala, G.~Smirnov and V.~Vlachoudis, Nucl. Data Sheets \textbf{120}, 211 (2014). 

\bibitem{Ferrari:2005zk}
  A.~Ferrari, P.~R.~Sala, A.~Fasso and J.~Ranft, \textbf{CERN-2005-10}, (2005).

\bibitem{Wibig:2021pim}
T.~Wibig, Chin. Phys. C \textbf{45}, 085001 (2021).

\bibitem{Engel:2018akg}
  R.~Engel, D.~Heck, T.~Huege, T.~Pierog, M.~Reininghaus, F.~Riehn, R.~Ulrich, M.~Unger and D.~Veberi\v{c}, Comput. Softw. Big Sci. \textbf{3} no.1,2 (2019).

%%%%%%%%%%%%%%%%%%%%%%%%%%%%%%%%%%%%%%%%%%%%%%%%%%%%%%%%

\bibitem{Barber:2017rvr} A.~S.~Barber, D.~B.~Kieda, Springer, R. W., Proceedings, 35th International Cosmic Ray Conference \textbf{ICRC2017}, (2017).

\bibitem{Mueller:2015thh} S.~Mueller and M.~Roth, PoS, \textbf{ICRC2015}, 419 (2016).  
  
\bibitem{Wentz:2003bp} J.~Wentz, I.~M.~Brancus, A.~Bercuci, D.~Heck, J.~Oehlschlager, H.~Rebel and B.~Vulpescu, Phys. Rev. D \textbf{67}, 073020, (2003).

\bibitem{Djemil:2005hr}
T.~Djemil, R.~Attallah and J.~N.~Capdevielle, Int. J. Mod. Phys. A \textbf{20} 6950 (2005).

\bibitem{Nikolaenko:2021oyi}
R.~V.~Nikolaenko, A.~G.~Bogdanov, R.~P.~Kokoulin and A.~A.~Petrukhin, Phys. At. Nucl. \textbf{84}, 1011 (2021).


\bibitem{Halataei:2008zz}
S.~M.~H.~Halataei, M.~Bahmanabadi, M.~K.~Ghomi and J.~Samimi, Phys. Rev. D \textbf{77}, 083001 (2008).  

\bibitem{Patgiri:2016odq}
P.~Patgiri, D.~Kalita and K.~Boruah, Indian J. Phys. \textbf{91}, 351 (2017). 



\bibitem{Bahmanabadi:2018dbr}
M.~Bahmanabadi and L.~Rafezi, Phys. Rev. D \textbf{98}, 103003 (2018).

\bibitem{Bahmanabadi:2019wdx}
M.~Bahmanabadi, Nucl. Instrum. Meth. A \textbf{945}, 162635 (2019).

\bibitem{Bahmanabadi:2019kel}
M.~Bahmanabadi and M.~Fazlalizadeh, Phys. Rev. D \textbf{100}, 083004 (2019).

\bibitem{Cohu:2021tjh}
A.~Cohu, M.~Tramontini, A.~Chevalier, J.~Marteau and J.~C.~Ianigro, JAIS \textbf{2022}, 250 (2022).



%\bibitem{Wibig:2021qkp}
%T.~Wibig, J. Phys. G \textbf{49}, 035201 (2022).

%\bibitem{Hariharan:2019xgn}
%B.~Hariharan, S.~R.~Dugad, S.~K.~Gupta, Y.~Hayashi, S.~S.~R.~Inbanathan, P.~Jagadeesan, A.~Jain, S.~Kawakami, P.~K.~Mohanty and B.~S.~Rao, Exper. Astron. \textbf{48}, 111 (2019).


\bibitem{shukla2016energy} 
P.~Shukla and S.~Sankrith, Int. J. Mod. Phys. A \textbf{33}, 1850175 (2018).

\bibitem{gaisser2002semi} Gaisser, T. K., Astropart. Phys., \textbf{16}, 285 (2002).

\bibitem{abe2016measurements} K.~Abe, {\it et. al}, Astrophys. J.\textbf{822}, 65 (2016).

\bibitem{adriani2011} O.~Adriani et al. (PAMELA), Science\textbf{332}, 69 (2011).

\bibitem{aguilar12015} M.~Aguilar et al. (AMS), Phys. Rev. Lett.\textbf{114}, 171103 (2015).


\bibitem{asakimori1998} K.~Asakimori et al., Astrophys. J. \textbf{502}, 278 (1998).

\bibitem{panov2009} A.D.~Panov et al. (ATIC Collab.), Bull. Russian Acad. of Science, Physics, \textbf{73}, 564 (2009).

\bibitem{derbina2005} V.A.~Derbina, {\it et. al} (RUNJOB), Astrophys. J. 628, L41 (2005).

\bibitem{aguilar22015} M.~Aguilar et al. (AMS), Phys. Rev. Lett.\textbf{115}, 211101 (2015).

\bibitem{Pierog:2015EPOSatLHC}  T.~Pierog \textit{et al.}, Phys. Rev. C \textbf{92}, 034906 (2015).

\bibitem{Graziani2017LHCb} G.~Graziani, PoS, \textbf{ICRC2017}, 214 (2017).

\bibitem{cecchini2012atmospheric} S.~Cecchini, and M.~Spurio, 2012, arXiv:1208.1171 (astro-ph).

\bibitem{crookes1972investigation} J.~N.~Crookes, and B.~C.~Rastin, Nucl. Phys. B \textbf{39}, 493 (1972).
    
\bibitem{dmitrieva2006measurements} A.~N.~Dmitrieva, {\it et. al}, Phys. Atom. Nucl. \textbf{69}, 865 (2006).

\bibitem{flint1972variation}  R.~W.~Flint, and R.~B.~Hicks, and S.~Standil Can. J. Phys. \textbf{50}, 843 (1972).

\bibitem{gettert1993} M.~Gettert, and J.~Unger and R.~Trezeciak, and J.~Engler and J.~Knapp, Proc. 23rd International Cosmic Ray Conference (ICRC), Calgary, 4, 394 (1993). 

\bibitem{haino2004measurements} Haino, S. {\it et. al}, Phys. Lett. B \textbf{594}, 35 (2004).
  
\bibitem{rastin1984accurate} B.~C.~Rastin, J. Phys. G \textbf{10}, 1609 (1984).

\bibitem{jokisch1979cosmic} H.~Jokisch, K.~Carstensen, W.~D.~Dau, H.~J.~Meyer and O.~C.~Allkofer, Phys. Rev. D \textbf{19}, 1368 (1979).
   
\bibitem{CMS:2010yju}
V.~Khachatryan \textit{et al.} [CMS], Phys. Lett. B \textbf{692}, 83 (2010).

\bibitem{MINOS:2007laz}
P.~Adamson \textit{et al.} [MINOS], Phys. Rev. D \textbf{76}, 052003 (2007).

\bibitem{OPERA:2010cos}
N.~Agafonova \textit{et al.} [OPERA], Eur. Phys. J. C \textbf{67}, 25 (2010).

\bibitem{L3:2004sed}
P.~Achard \textit{et al.} [L3], Phys. Lett. B \textbf{598}, 15 (2004).

\end{thebibliography}
\end{document}